\documentclass[superscriptaddress,twocolumn,preprintnumbers,amsmath,amssymb,aps]{revtex4}
\pdfoutput=1

\usepackage{amsmath,amssymb,amsthm,amsfonts}
\usepackage{MnSymbol}
\usepackage{mathrsfs}
\usepackage{slashed}
\usepackage{graphicx}
\usepackage{subfigure}
\usepackage{color,rotating}
\usepackage{dsfont}
\usepackage{setspace}
\usepackage{verbatim}
\usepackage{hyperref}
\usepackage[export]{adjustbox}
\usepackage[normalem]{ulem}

\makeatletter
\def\l@subsubsection#1#2{}
\makeatother

\newcommand{\beq}{\begin{equation}}
\newcommand{\eeq}{\end{equation}}
\newcommand{\beqa}{\begin{eqnarray}}
\newcommand{\eeqa}{\end{eqnarray}}
\newcommand{\bfc}{\begin{figure}[t]\begin{center}}
\newcommand{\efc}{\end{center}\end{figure}}

\def\Fig#1{Fig.~\ref{#1}}
\def\fig#1{Fig.~\ref{#1}}
\def\Eq#1{Eq.~(\ref{#1})}
\def\eq#1{(\ref{#1})}

\def\0#1#2{\frac{#1}{#2}}  

\def\CC{{\mathcal C}}

\def\CK{{\mathcal K}}

\newcommand{\Tr}{\mathrm{Tr}}

\newcommand{\be}{\begin{eqnarray}}
\newcommand{\ee}{\end{eqnarray}}

\def\flow{{\CK}}

\def\ip{p\llap{/}}

\begin{document}

\title{Physics and the choice of regulators in functional renormalisation group flows}

\author{Jan M. Pawlowski} \affiliation{Institut f\"ur Theoretische
  Physik, Universit\"at Heidelberg, Philosophenweg 16, 69120
  Heidelberg, Germany} \affiliation{ExtreMe Matter Institute EMMI, GSI
  Helmholtzzentrum f\"ur Schwerionenforschung mbH, Planckstr. 1,
  D-64291 Darmstadt, Germany}

\author{Michael M. Scherer} \affiliation{Institut f\"ur Theoretische
  Physik, Universit\"at Heidelberg, Philosophenweg 16, 69120
  Heidelberg, Germany}

\author{Richard Schmidt} \affiliation{ITAMP, Harvard-Smithsonian
  Center for Astrophysics, 60 Garden Street, Cambridge, MA 02138, USA}
\affiliation{Physics Department, Harvard University, 17 Oxford Street,
  Cambridge, MA 02138, USA}

\author{Sebastian J. Wetzel} \affiliation{Institut f\"ur Theoretische
  Physik, Universit\"at Heidelberg, Philosophenweg 16, 69120
  Heidelberg, Germany}


\begin{abstract}
  The Renormalisation Group is a versatile tool for the study of many
  systems where scale-dependent behaviour is important.  Its
  functional formulation can be cast into the form of an exact flow
  equation for the scale-dependent effective action in the presence of
  an infrared regularisation. The functional RG flow for the
  scale-dependent effective action depends explicitly on the choice of
  regulator, while the physics does not. In this work, we
  systematically investigate three key aspects of how the regulator
  choice affects RG flows: (i) We study flow trajectories along closed
  loops in the space of action functionals varying both, the regulator
  scale and shape function. Such a flow does not vanish in the
  presence of truncations.  Based on a definition of the length of an
  RG trajectory, we suggest a practical procedure for devising
  optimised regularisation schemes within a truncation.  (ii) In
  systems with various field variables, a choice of relative cutoff
  scales is required.  At the example of relativistic bosonic
  two-field models, we study the impact of this choice as well as its
  truncation dependence.  We show that a crossover between different
  universality classes can be induced and conclude that the relative
  cutoff scale has to be chosen carefully for a reliable description
  of a physical system.  (iii) Non-relativistic continuum models of
  coupled fermionic and bosonic fields exhibit also dependencies on
  relative cutoff scales and regulator shapes. At the example of the
  Fermi polaron problem in three spatial dimensions, we illustrate
  such dependencies and show how they can be
  interpreted in physical terms.
\end{abstract}

\maketitle


\section{Introduction}

\noindent In the past twenty years, the functional renormalisation
group (FRG) approach \cite{Wetterich:1992yh} has been established as a
versatile method allowing to describe many aspects of different
physical systems in the framework of quantum field theory and
statistical physics.  Applications range from quantum dots and wires,
statistical models, condensed matter systems in solid state physics
and cold atoms over quantum chromodynamics to the standard model of
particle physics and even quantum gravity. For reviews on the various
aspects of the functional RG see
\cite{Litim:1998nf,Aoki:2000wm,Bagnuls:2000ae,Berges:2000ew,Polonyi:2001se,%
  Pawlowski:2005xe,Gies:2006wv,
  Schaefer:2006sr,Delamotte:2007pf,Kopietz:2010zz,Scherer:2010sv,Rosten:2010vm,
  Braun:2011pp,Litim:2011cp,Percacci:2011fr,Boettcher:2012cm,%
vonSmekal:2012vx,Reuter:2012id}.

The functional renormalisation group approach can be set-up in terms
of an exact flow equation for the effective action of the given theory
or model \cite{Wetterich:1992yh}. The choice of the initial condition
at some large ultraviolet cutoff scale, typically a high momentum or
energy scale, together with that of the regulator function determines
both, the physics situation under investigation as well as the
regularisation scheme. The functional RG flow for the scale-dependent
effective action depends explicitly on the choice of regulator, while
the physics does not. The latter is extracted from the full quantum
effective action at vanishing cutoff scale implying a vanishing
regulator. Hence, at this point no dependence on the choice of
regulator is left, only the implicit choice of the regularisation
scheme remains.

Typically, for the solution of the functional flow equation for the
effective action one has to resort to approximations to the effective
action as well as to the flow. Such a truncation of the full flow
usually destroys the regulator independence of the full quantum
effective action at vanishing cutoff. Therefore, devising suitable
expansion schemes and regulators is essential for 
reliable results. Moreover, the related considerations also allow for a
discussion of the systematic error within a given truncation
scheme. This has been examined in detail for the computation of
critical exponents in models with a single scalar field in three
dimensions within the lowest order of the local potential
approximation (LPA),
\cite{Ball:1994ji,Liao:1999sh,Litim:2000ci,Litim:2001up,%
  Litim:2001fd,Litim:2002cf,Litim:2002qn,Canet:2002gs,%
  Litim:2005us,Pawlowski:2005xe}: an optimisation procedure,
\cite{Litim:2000ci,Pawlowski:2005xe} suggests a particular regulator
choice -- the flat regulator -- which is also shown to yield the best
results for the critical exponents.

The optimisation framework in \cite{Pawlowski:2005xe} has been
extended to general expansion schemes in a functional optimisation
procedure including fully momentum-dependent approximation schemes. An
application to momentum-dependent correlation functions in Yang-Mills
theory can be found in \cite{Fischer:2008uz}.

Still, for more elaborate truncations, in particular higher orders of
the derivative expansion in the LPA, including, e.g., momentum
dependencies or higher-order derivative terms, little has been done
when it comes to a practical implementation of optimisation
criteria. Also, more complex physical models with different symmetries
such as, e.g., non-relativistic systems, or models with several
different fields, for example mixed boson-fermion systems, demand for
a thorough study of their regulator dependence in order to extract the
best physical results from a given truncation. 

In this paper we study the impact of different regulator choices on
truncated functional renormalisation group flows in various models and
further develop the functional optimisation procedure set-up in
\cite{Pawlowski:2005xe}. In its present form it allows for a practical
and simple comparison of the quality of different regulators and for
the construction of an optimised one.  

The rest of the paper is organised as follows: In
Sec. \ref{sec:frgflows} we shortly introduce the FRG approach and
explain how the choice of a specific regulator influences a truncated
FRG flow.  This is captured in terms of an integrability condition for
closed loops in theory space upon a change of the regulator and RG
scale, cf. Sec. \ref{sec:scalarloops}.  In Sec. \ref{sec:opt} we then
devise a road towards a practical optimisation procedure.  We discuss
the length of an RG trajectory which has to be minimal for an optimised
regulator, cf. Sec. \ref{sec:RGlength}. This procedure is then applied
to a simply scalar model (Sec.~\ref{sec:pracopt}).  A more heuristic
approach to models with various degrees of freedom such as two-scalar
models and non-relativistic boson-fermion systems, is presented in
Secs. \ref{sec:multifield} and \ref{sec:richard}, respectively.  To
this end, we introduce a shift between the regulator scales of the
different fields and show how this affects the results allowing for a
change of the underlying physics upon varying the regulator.  This,
again, clearly demands for carefully choosing a regularisation scheme
which could be performed by a optimisation procedure as suggested in
this work.


\section{Functional RG Flows}\label{sec:frgflows}

\noindent The functional renormalisation group is based on the
Wilsonian idea of integrating out degrees of freedom.  In the
continuum, this idea can be implemented by suppressing the
fluctuations in the theory below an infrared cutoff scale $k$.  An
infinitesimal change of $k$ is then described in terms of a
differential equation for the generating functional of the theory at
hand -- Wetterich's flow equation \cite{Wetterich:1992yh}.  The
infrared suppression can be achieved by adding a momentum-dependent
mass term to the classical action,
\begin{align}\label{eq:Sk}
S[\varphi] \to S[\varphi] + \012 \int_p \varphi(p) R_k(p) \varphi(-p)\,, 
\end{align}
with $\int_p=\int \0{d^d p}{(2 \pi
  )^d}\,.$ The regulator $R_k(p)$ tends towards a mass for low momenta
and vanishes sufficiently fast in the ultraviolet, see \eq{eq:r} and
\eq{eq:rlims}.

With the cutoff scale dependent action \eq{eq:Sk} also the
one-particle-irreducible (1PI) effective action or free energy,
$\Gamma_k[\phi]$, acquires a scale dependence.  The $n$th field derivatives of
the effective action, $\Gamma_k^{(n)}[\phi]$, are the 1PI parts of the
$n$-point correlation functions in a general background $\phi=\langle
\varphi\rangle$.  The flow of $\Gamma_k$ is given by
\begin{align}
  \partial_t {\Gamma}_k&=\frac{1}{2} \textrm{Tr}\, G_k[\phi]\,
  \partial_t {R}_k\ {\rm with}\ G_k[\phi]=
  \frac{1}{\Gamma_k^{(2)}[\phi]+R_k}\label{eq:flow}\,,
\end{align}
where we have introduced the renormalisation time $t= \ln k/\Lambda$. 
Here, $\Lambda$ is some reference scale, usually the ultraviolet
scale, where the flow is initiated.  The trace sums over all occurring
indices, including the loop integration over momenta.
The regulator is conveniently written as
\begin{align} \label{eq:r}
R_k(p) = p^2\, r( p^2/k^2 )\,, 
\end{align}
with the dimensionless shape function $r(y)$ that only depends on
the dimensionless ratio $y=p^2/k^2$.
The regulator functions fulfill the infrared and ultraviolet conditions
\begin{align}\label{eq:rlims} 
  \lim_{y\rightarrow 0} y\, r(y) > 0 \,,\quad
  \lim_{y\rightarrow\infty, \epsilon >0}y^{d/2+\epsilon}\, r(y) =0\,.
\end{align}
The first limit in \eq{eq:rlims} implements the infrared suppression
of low momentum modes as the propagator $G_k$ acquires an additional
infrared mass due to $R_k$.  The second limit guarantees that the
ultraviolet is unchanged.  The regulator $R_k(p)$ has to decay with
higher powers as $p^d$ in $d$ dimensions in order to have a
well-defined flow equation without the need of an ultraviolet
renormalisation.  With~\eq{eq:rlims} the flow equation \eq{eq:flow} is
ultraviolet finite due to the sufficiently fast decay of the regulator
in the ultraviolet. Here, we presented the relativistic case for
simplicity.  The arguments can be extended to the non-relativistic
case, e.g.~\cite{Diehl:2007ri}. We discuss one specific example for
such a non-relativistic system in Sec.~\ref{sec:richard}.


\subsection{Ultraviolet Limit and Regulator Dependence}\label{sec.uvreg}

\begin{figure*}[t!]
\includegraphics[width=\textwidth]{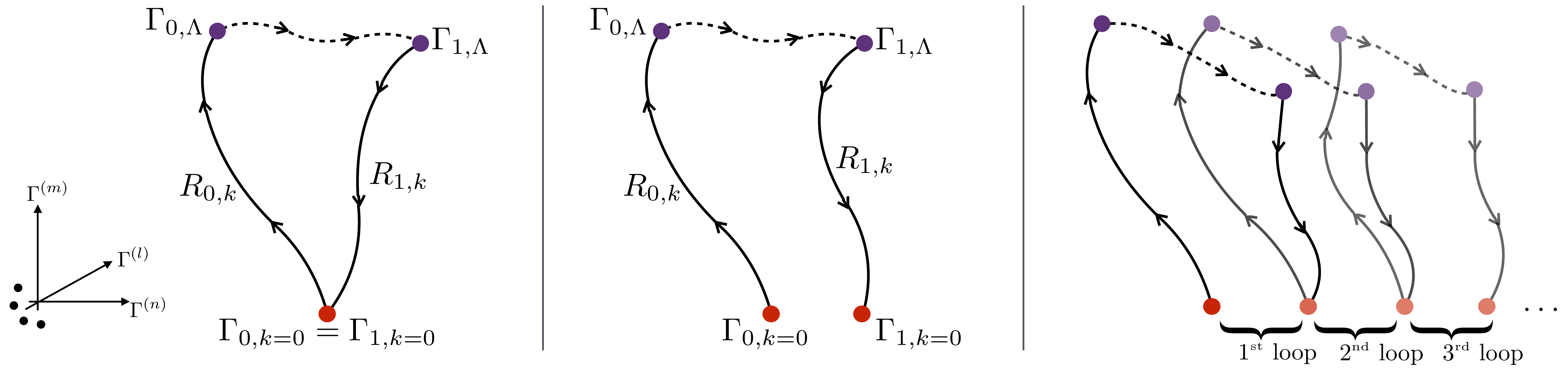}\\
\caption{Left panel: Pictorial representation of the integrability
  condition \eq{eq:intcond} in the theory space of action
  functionals. By means of \eq{eq:intflow}, we can map the two actions
  at the initial scale $\Lambda$ onto each other. Integrating out
  momentum fluctuations from $k=\Lambda$ toward $k=0$ then yields the
  full quantum effective action for both regulators, i.e.,
  $\Gamma_{0,k=0}=\Gamma_{1,k=0}$. Middle panel: Approximations lead
  to a violation of the integrability condition,
  cf.~\eq{eq:closedloop}. Right panel: Pictorial representation of
  several successively performed loops according to \eq{eq:closedloop}
  within an approximation for the full propagator $G_k$. This sketch
  exhibits how the result for the full effective action $\Gamma_{k=0}$
  moves further and further away from its initial position in theory
  space.}
\label{fig:globalloop}
\end{figure*}

\noindent In the limit $k\to \infty$, the cutoff term in \eq{eq:Sk}
suppresses all momentum fluctuations.  To discuss this limit, we
consider the RG running of the scale-$k$-dependent couplings
$g_{n}(k)$ parametrising the theory in terms of a suitable basis of
field monomials, i.e., $\Gamma_k=\sum_n
g_n(k)\,\mathcal{O}_n(\partial, \phi)\,.$ We classify the $g_{n}(k)$
according to their UV scaling dimension $d_{n}$ that follows from the
running of the couplings towards the ultraviolet (UV) with the flow
equation~\eq{eq:flow}. The UV scaling dimension $d_n$ is the full
quantum dimension, i.e., canonical plus anomalous dimension,
\begin{align}
	g_n(k) \sim k^{d_{n}}\,.
\end{align}
Terms in the effective action $\Gamma_k$ whose couplings $g_n(k)$ have
semi-positive UV scaling dimension, $d_{n}\geq 0$, dominate the UV behaviour.
In turn, terms with couplings $g_n(k)$ with $d_{n}< 0$ are
sub-leading or suppressed.

Let us elucidate this at the example of the relativistic $\varphi^4$
field theory in $d=3$ dimensions.  This theory is super-renormalisable
and the only parameter with a positive UV scaling dimension is the
mass parameter $m^2_k=\Gamma_k^{(2)}(p=0,\phi=0)$.  The flow of the
mass is derived from \eq{eq:flow} with a second order field derivative
evaluated at vanishing fields and momenta, to wit
\begin{align}\label{eq:flowm}
  \partial_t m_k^2 = -\012 \int_q G_k(q)\,\partial_t R_k(q)\,
  G_k(q)\,\Gamma_k^{(4)}(q,q,0,0)\,, 
\end{align}
where we have used that the three-point function vanishes due to the
symmetry of the theory under $\phi\to -\phi$, i.e.,
$\Gamma_k^{(3)}[\phi=0]=0$. In the UV limit the flow \eq{eq:flowm}
simplifies considerably. The four-point function tends towards a local
scale-independent vertex, i.e.,
$\Gamma_{k\rightarrow\infty}^{(4)}\rightarrow\lambda_{\text{\tiny
    UV}}$, up to momentum conservation.  For $k\to \infty$, the
propagators in~\eq{eq:flowm} are simply given by
\begin{align}\label{eq:prop3duv} 
G_k(q) = \0{1}{q^2[1+r(q^2/k^2)]   + m_k^2}\,.
\end{align}
Here, we have also used that the wave function renormalisation $Z_k\to
1 +O(1/k)$ in the limit $k\to \infty$. This can be proven analogously
to the following determination of the asymptotic scaling of the
mass. For the mass we are hence led to the asymptotic UV flow
\begin{align}\label{eq:flowmUV}
\partial_t m_k^2 = - k\, \lambda_{\text{\tiny UV}} 
\int_{\bar q} \0{\bar q^4 \partial_{\bar q^2} r(\bar q^2) }{( \bar q^2[1+r(\bar q^2)] +
    \bar m_k^2)^2}  \,, 
\end{align}
where quantities denoted with a bar are scaled with appropriated
powers of $k$ in order to make them dimensionless, i.e., $\bar
q^2=q^2/k^2$ and $\bar m_k^2=m_k^2/k^2$.  The flow \eq{eq:flowmUV} is
further simplified if we reduce it to the leading UV scaling.  To that
end, we notice that the flow scales with $k$ for $\bar m_k=0$.  Hence,
for $k\to \infty$ we have $m_k^2 \propto \lambda_{\text{\tiny UV}}\,
k$ and $\bar m_k^2 \propto \lambda_{\text{\tiny UV}}/k\to 0$.
Accordingly, we have
\begin{align}\nonumber 
  m_k^2 = &\mu(r) \lambda_{\text{\tiny UV}}\,
  k+O(k^0)\,,\label{eq:mUV}
\end{align}
with the dimensionless factor 
\begin{align}
 \mu(r) = &- \int_{\bar q} \0{\partial_{\bar q^2}
    r(\bar q^2) }{[1+r(\bar q^2)]^2} \,. 
\end{align}
We conclude that the mass parameter $m_k^2$ diverges linearly with
$k$. We also note that the constant $\mu(r)$ is non-universal and
depends on the chosen regulator $r$. Additionally, the above simple
example nicely reflects the regularisation and renormalisation scheme
dependence in the present modern functional RG setting: UV divergences
in standard perturbation theory are reflected in UV relevant terms
such as $\mu(r) \lambda_{\text{\tiny UV}}\, k$, that diverge for
$k\to\infty$. The subtractions or renormalisation in perturbation
theory are reflected in the consistent choice of the initial condition
that makes the full effective action $\Gamma_{k=0}$ independent of the
initial scale $k=\Lambda$. Accordingly, the initial mass $m^2_\Lambda$
has to satisfy the flow equation \eq{eq:flowm} which again leads to
\eq{eq:mUV} for $m_\Lambda^2$. In other words, the
$\Lambda$-dependence of the flow is annihilated by that of the initial
conditions. This accounts for a BPHZ-type renormalisation, for
detailed discussions see
e.g.~\cite{Litim:2002xm,Pawlowski:2005xe,Rosten:2010vm}. Consequently,
a part of the standard renormalisation scheme dependence is carried by
the regulator dependence of~$\mu(r)$.
 
Finally, the physics is entirely carried by the finite part of the
UV limit, that is the $O(k^0)$ term in \eq{eq:mUV}.  Since this finite
UV part has first to be mapped to $k\to0$ via the flow, it also
carries a renormalisation scheme dependence. In summary, the latter is
given by a combination of the shape dependence and the finite part of
the initial condition. This simple distinction can be used to rewrite
the effective action in terms of renormalised fields and parameters
for obtaining a finite UV limit, see
e.g.~\cite{Pawlowski:2005xe,Rosten:2010vm}.


\subsection{Initial Actions and Integrability Condition}\label{sec:init}

\noindent The above discussion already highlights the regulator- or
$r$-dependence of the flow and the scale-dependent effective
action. However, despite this $r$-dependence of the flow, the final
effective action $\Gamma_{k=0}$ is unique up to RG transformations,
see App.~\ref{app:rgtrafo} for a discussion of this issue.  This is
illustrated in theory space in the left panel of \fig{fig:globalloop}.
The initial effective actions at the UV scale $\Lambda$ differ due to
the different shape functions $r$.  Nonetheless, we can map the
initial effective actions onto each other by the following flow
equation
\begin{align} \label{eq:flow12} 
  \partial_s \Gamma_{s,\Lambda} = \frac{1}{2} \textrm{Tr}\,
  G_\Lambda[\phi]\, \partial_s R_{s,\Lambda}\,,
\end{align}
with $R_{s,k}= p^2 r_s$ and where we use a one parameter family of
shape functions $r_s(y)$ with $r_0(y) = r^{\text{A}}(y)$ and $r_1(y) =
r^{\text{B}}(y)$ which analytically transforms $r^{\text{A}}$ into
$r^{\text{B}}$. Then, \eq{eq:flow12} is easily derived similarly to
the cutoff flow \eq{eq:flow}. Consequently, the two initial actions
are mapped onto each other by the $s$-integration of \eq{eq:flow12},
\begin{align} \label{eq:intflow} \Gamma_{1,\Lambda} =
  \Gamma_{0,\Lambda} +\012 \int_0^1 ds\, \textrm{Tr}\,
  G_{\Lambda}[\phi]\, \partial_s R_{s,\Lambda}\,,
\end{align}
and the full quantum effective actions agree trivially,
$\Gamma_{0,k=0}=\Gamma_{1,k=0}$, see also left panel of
\fig{fig:globalloop}.
This statement can be reformulated as an integrability condition
\begin{align} \label{eq:intcond} \int_0^1 \partial_s
  \Gamma_{s,\Lambda} + \int_\Lambda^0 \0{dk}{k}\partial_t \Gamma_{1,k}
  +\int_0^\Lambda \0{dk}{k}\partial_t \Gamma_{0,k} =0\,,
\end{align}
which defines a closed loop in theory space.


\subsection{Integrability Condition and Approximations}
\begin{center}
\begin{figure}[t!]
\includegraphics[height=0.16\textheight]{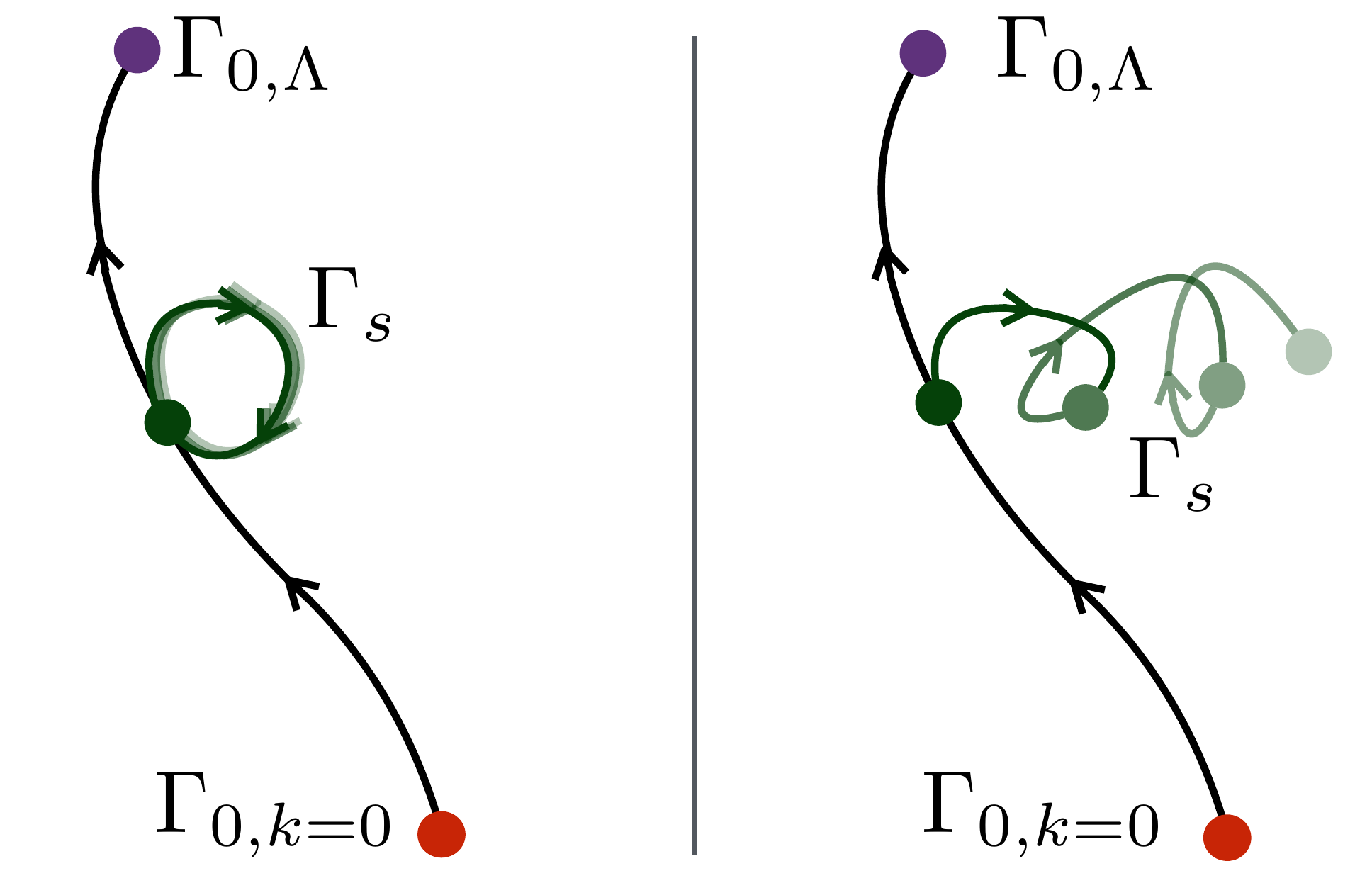}\\
\caption{Generalized one-parameter flows. Left panel: The exact flow
  equation provides the integrability condition,
  \eq{eq:closedloop}. Right panel: In general, the integrability condition is
  violated when approximations to $G_k$ are employed,
  cf. \eq{eq:closedloop2}.}
\label{fig:sflow}
\end{figure}
\end{center}
\noindent In general, the integrability condition \eq{eq:intcond} is
violated in approximations to the full effective action, and the flow
ceases to describe a total derivative with respect to $t$ and $s$. The
relation between an approximation of the effective action or
rather of the propagator in the flow equation and the derivative
operators $\partial_t$ and $\partial_s$ is more clearly seen in the
general flow equation for composite operators $I_k$. This set includes
general $n$-point correlation functions $ \langle \phi(x_1)\cdots
\phi(x_n)\rangle$ with connected and disconnected
parts~\cite{Pawlowski:2005xe}. A further relevant example is
$\delta\Gamma_k/\delta\phi$. The flow equation for the $I_k$ reads
\begin{align}\label{eq:flowcomp} 
  \partial_t I_k[\phi] = \left( -\012 \Tr\, G_k[\phi]\,\partial_t R_k
    \, G_k[\phi]\0{\delta^2}{\delta\phi^2} \right) \,  I_k\,.
\end{align}
For $\delta\Gamma_k/\delta\phi$ one easily sees that the
$\phi$-derivative of the flow \eq{eq:flow} gives \eq{eq:flowcomp} with
$I_k=\Gamma_k^{(1)}$. Note, however, that the effective action
$\Gamma_k$ does not satisfy \eq{eq:flowcomp}. Another simple test is
given by the full two-point function
$G_k(p,q)+\phi(p)\phi(q)$. Importantly, as the set of composite
operators that satisfy \eq{eq:flowcomp} includes all correlation
functions it is complete. We conclude that the total $t$-derivative
has the representation 
\begin{align}\label{eq:repdt}
  \partial_t = \left( -\012 \Tr\, G_k[\phi]\,\partial_t R_k\,
    G_k[\phi]\0{\delta^2}{\delta\phi^2} \right) \,, 
\end{align}
on the -- complete -- set of composite operators $\{I_k\}$. \Eq{eq:repdt}
makes explicit the consequences of approximations to the effective
action for the total $t$-derivative: they imply approximations for the
full propagators $G_k$ and hence approximations to the representation
\eq{eq:repdt} of the total $t$-derivative $\partial_t$. Consequently,
an integration along a closed loop in regulator space does not
necessarily vanish within an approximation to the effective
action. Note that the notation $\partial_t$ for the total
$t$-derivative is common in the FRG community and we keep it for the
sake of comparability.

For our discussion of flows that change regulators as well as cutoff
scales we extend the notation with the parameter $s$ to general
one-parameter flows in theory space.  Such a flow includes changes of
the cutoff scale $k$ with $k(s)$, changes of the shape of the
regulator $r_s$ and reparametrisations of the theory, see
App.~\ref{app:onparam} for a detailed discussion.  Within this unified
approach a closed loop such as the global one in \eq{eq:intcond} has
the simple representation
\begin{align}\label{eq:closedloop}
  \oint_\CC d\, \Gamma[\phi,R] = \int_0^1
  ds\,\frac{d}{ds}\Gamma_s[\phi,R_s] =0\,,
\end{align}
Here, $s$ parameterises the closed curve $\CC$, and $R_s$ describes a
loop in regulator space with $R_{0}=R_{1}$.  In general,
approximations to the effective action $\Gamma_k$ lead to
\begin{align}\label{eq:closedloop2}
 \int_0^1 ds\,\frac{d}{ds}\Gamma_s[\phi,R_s] \neq 0\,,
\end{align}
for closed loops, see Fig.~\ref{fig:sflow} for a pictorial
representation.  This also means that if such a loop is taken many
periods eventually the result will be dominated completely by the
errors introduced by the approximation of the representation to the
total $t$-derivative.  In particular, the global loop shown in the
left panel of \fig{fig:globalloop} does not close. 

\begin{center}
\begin{figure}[t!]
\includegraphics[height=0.22\textheight]{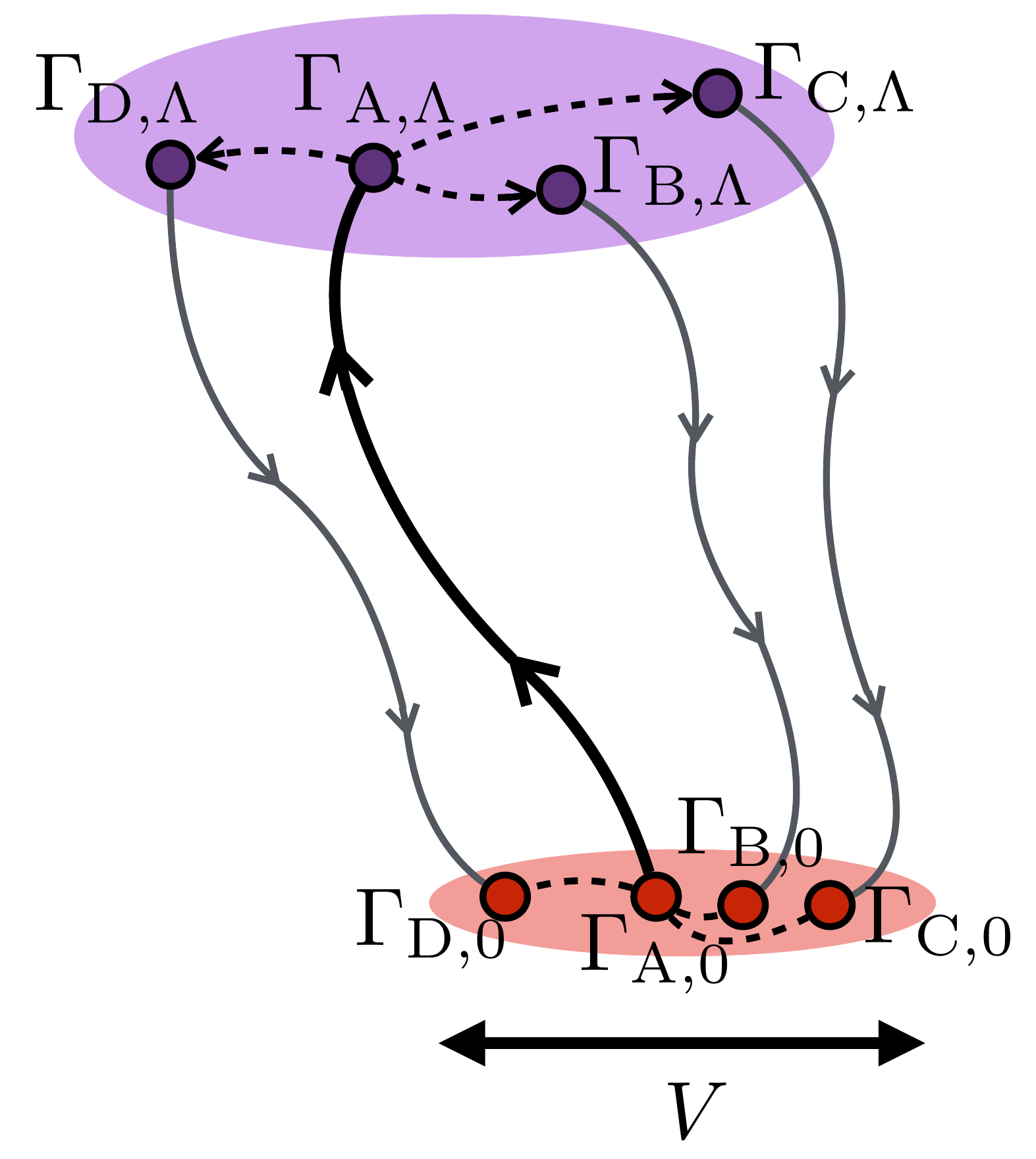}\\
\caption{A change of regulator at the UV scale $\Lambda$,
  \Eq{eq:intflow}, maps the action $\Gamma_{\text{A},\Lambda}$ onto
  one of the others, $\Gamma_{\text{B},\Lambda},
  \Gamma_{\text{C},\Lambda}, \Gamma_{\text{D},\Lambda}$, corresponding
  to different regulators. Then, integrating the RG flow in an
  approximation towards $k=0$ results in different approximations to
  the IR quantum effective action, i.e., $\Gamma_{\text{i},0}$ with
  $i\in \{\text{A,B,C,D}\}$, cf. \Eq{eq:closedloop}. The spreading of
  these results from a large class of general regulators can be used
  as an error estimate for an approximation.}
\label{fig:spread}
\end{figure}
\end{center}

A violation \eq{eq:closedloop2} of the integrability condition
\eq{eq:closedloop} is a measure for the self-consistency of the
approximation at hand. In the following we will use it in our quest
for optimal regulators as well as a systematic error estimate.  In
App.~\ref{app:intcond} we discuss under which
circumstances~\eq{eq:closedloop} is violated and when it is satisfied.
In summary, the violation of the integrability condition measures the
incompleteness, in terms of the full quantum theory, of fully
non-perturbative resummation schemes.  This allows for a systematic
error estimate of a given approximation: Consider general closed loops
in theory space initiated from a given regulator
$R^{\text{A}}_k$. Then, we change the regulator at a fixed initial
scale as in~\eq{eq:flow12}, and subsequently flow to vanishing cutoff
scale. For sensible regulator choices, the spreading of the results
for $\Gamma[\phi,R=0]$ provides an error estimate,
see~\fig{fig:spread} for a pictorial representation.
%


\subsection{Scalar model}\label{sec:scalarloops}

\begin{table*}[t!]
  \setlength{\tabcolsep}{4pt}
  \renewcommand{\arraystretch}{1.5}
  \begin{tabular}{c|c|c}\hline\hline
    Regulator type & Representation & Limits \\ \hline \hline
    exponential & $r_{\text{exp}}(y)=(\exp(y)-1)^{-1}$ & - \\ \hline
    flat (Litim) & $r_{\text{flat}}(y)=(\frac{1}{y}-1)\theta(1-y)$ & - \\ \hline 
    step-like & $r_{\text{sl}}(y)=\frac{c}{y}\theta(1-y)$ & 
    $\lim_{c\rightarrow\infty}r_{\text{sl}}(y)\rightarrow r_{\text{sharp}}(y)$\\ \hline
    compactly supported smooth& $r_{\text{css}}(y)=
    \frac{c\,\theta(1-hy^b)}{\exp({\frac{cy^b}{1-hy^b})-1}}$ 
    & $\begin{matrix} \text{(i)}& \lim_{b\rightarrow\infty} r_{\text{css}}(y)
      \rightarrow r_{\text{sharp}}(y)\,,\ \text{for}\ c>0,h= 0\\
      \text{(ii)}& \lim_{c\rightarrow 0}r_{\text{css}}(y)\rightarrow
      r_{\text{flat}}(y)\,,\ \text{for}\ b=1, h= 1\\
      \text{(iii)}& r_{\text{css}}(y) =r_{\text{exp}}(y)\,,\ 
q\text{for}\ b=c=1, h=0
\end{matrix}$ \\ \hline exponential interpolating &
$r_{\text{int}}(y)=\frac{(a-by)y^{n-1}}{\exp(y^{n})-1}$
&$\begin{matrix} \text{(i)}& \lim_{a, n\rightarrow
    \infty}r_{\text{int}}(y)
  \rightarrow r_{\text{sharp}}(y)\,,\  \text{for}\ b=0\\
  \text{(ii)}& \lim_{n\rightarrow\infty}r_{\text{int}}(y)
  \rightarrow r_{\text{flat}}(y)\ , \text{for} \ a=1,b=1\\
  \text{(iii)}& r_{\text{int}}(y)=r_{\text{exp}}(y)\,,\ \text{for}\
  a=1, b=0, n=1
\end{matrix}$ \\
 \hline
 \hline
  \end{tabular}
  \caption{Classes of regulator shape functions $r$, and parameter choice that provide  
    exponential, flat and sharp regulators in the respective class.}\label{tregs}
\end{table*}
\noindent Generally, it is not possible to exactly solve the flow
equation~\eq{eq:flow} for the flowing action $\Gamma_k$.  Therefore,
we have to devise suitable truncation schemes for the functional
$\Gamma_k$. A simple scheme is given by the derivative expansion which
assumes the smallness of momentum fluctuations. In this section, we
will investigate a three-dimensional $O(1)$ symmetric scalar model to
explore the effects of regulator choices on functional RG results.
Our ansatz is given by the local potential approximation (LPA)
\begin{align}\label{eq:potexp}
  \Gamma_k=\int_x \Big[\frac{1}{2} (\partial_\mu\phi)^2
  +U_k(\phi)\Big]\,,\ \ U_k=\sum_i
  \frac{\bar\lambda_i}{i!}(\bar\rho-\bar\kappa)^i\ ,
\end{align}
with the real scalar field $\phi$, the scale-dependent effective
potential $U_k$ and the field invariant $\bar\rho=\frac{1}{2}\phi^2$.

The expansion parameters of the effective potential are scale
dependent quantities $\bar\lambda_n=\bar\lambda_n(k)$ and
$\bar\kappa=\bar\kappa(k)$, however for brevity, we will not indicate
this in the following.  Further, we have set the wave function
renormalisation to unity, dropping all non-trivial 
momentum-dependences. For calculations, we introduce the dimensionless
effective potential and couplings $u(\rho)=U_k k^{-d}$,
$\kappa=k^{2-d}\bar\kappa,\ \rho=k^{2-d}\bar\rho$ and
$\lambda_i=\bar\lambda_i k^{-d+i(d-2)}$.  Then, we can write the flow
equation for the effective potential as
\begin{align}
\partial_t u=-d\,u+(d-2)\rho\, u'+I(u'+2\rho\, u'')\,,
\end{align}
where primes denote derivatives with respect to $\rho$ and we have
defined the threshold function
\begin{align}\label{eq:uthresh}
  I(w)&=v_d\int_0^\infty y^{\frac{d}{2}+1}dy \
  \frac{-2\,\partial_yr(y)}{y(1+r(y))+w}\,,
\end{align}
with $y=p^2/k^2$ and $v_d^{-1}=2^{d+1}\pi^\frac{d}{2}\Gamma (d/2)$.
Using the series expansion of the effective potential, \eq{eq:potexp},
also for its dimensionless version, we extract the flow equations for
the individual couplings by projections, see
App.~\ref{app:twofieldcouplings} for explicit expressions.


\subsubsection{Switching Regulators at Fixed RG Scale}
\label{switching regulators}

\begin{figure}[t!]
  \centering
  \includegraphics[width=1.\columnwidth]{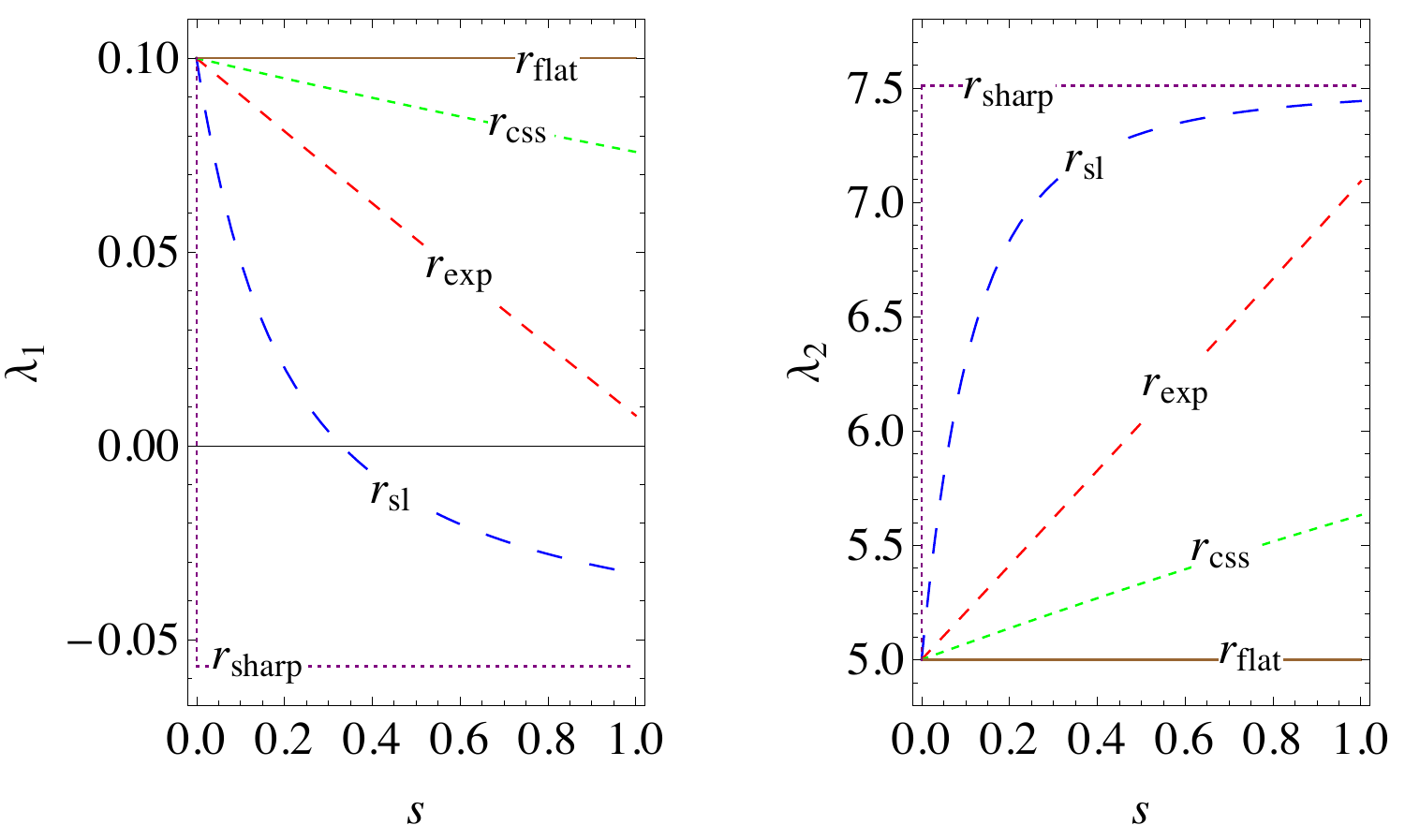}
  \caption{Change of the initial conditions for different choices of
    interpolations from $r_{\text{\tiny{flat}}}$ to $r_i$. The
    parameter-dependent regulator $r_{\text{\tiny{css}}}$ was chosen
    with $c=0.5, b=1, h=0.5$. The shape function
    $r_{\text{\tiny{sl}}}$ with $c=10$.}
\label{icond01}
\end{figure}

\noindent As discussed in Sect.~\ref{sec:init} a change in the
regulator triggers a flow in the space of action functionals.  In
particular this means that switching from one regulator to another
induces a change in the initial conditions as exhibited in
\eq{eq:intflow}.  To visualise this change explicitly, we employ
superpositions of two regulators at a fixed scale $k$, with $s \in
[0,1]$, 
\begin{align}
  r_{s}(y)=s\, r^{\text{A}}(y)+(1-s)\, r^{\text{B}}(y) \,.
\label{linsup}
\end{align}
The flow equation with respect to the variable $s$ is then, 
\begin{align}\nonumber 
  \partial_s  u&= J(u'+2\rho\, u'') \,,\\[2ex]
  J(w)&=v_d\int_0^\infty y^{\frac{d}{2}}dy
  \frac{\partial_s r_s(y)}{y(1+ r_s(y))+w}\,,
\label{dulinsup}
\end{align}
where $\partial_s r_s(y)=r^{\text{A}}(y)- r^{\text{B}}(y)$.  More
generally, we do not require a linear superposition as specified in
\eq{linsup}, but we can switch regulators on an arbitrary smooth
trajectory while keeping the scale $k$ fixed.  The change of initial
conditions from switching between different regulators is then given
by the solution of the flow equation \eq{dulinsup}.

In Fig.~\ref{icond01}, we show this solution for a collection of
representative regulator shape functions listed in Tab.~\ref{tregs}.
Here, we have integrated flow equations within an LPA to
fourth order in the fields $\phi^4$ in the symmetric regime,
concentrating on the four-scalar coupling and the only relevant
coupling, the mass parameter. In Fig.~\ref{icond01} we follow the
regulator-dependence of the two dimensionless couplings
$\lambda_1=m^2$ and $\lambda_2$, starting at $s=0$ with the flat
regulator $r_{\text{flat}}$ and the initial conditions
$\lambda_1^{\text{(in)}}=0.1$ and
$\lambda_2^{\text{(in)}}=5.0$. Fig.~\ref{icond01} clearly exhibits the
change in the initial conditions upon variations of the regulator
shape function at fixed RG scale $k$.  Interestingly, the largest
difference in initial conditions starting from
$r_{\text{flat}}$ is given by switching to the sharp regulator
$r_{\text{sharp}}$.


\subsubsection{Loops in $k-R_k$ Space}

\begin{figure}[t!]
  \centering
  \includegraphics[width=0.7\columnwidth]{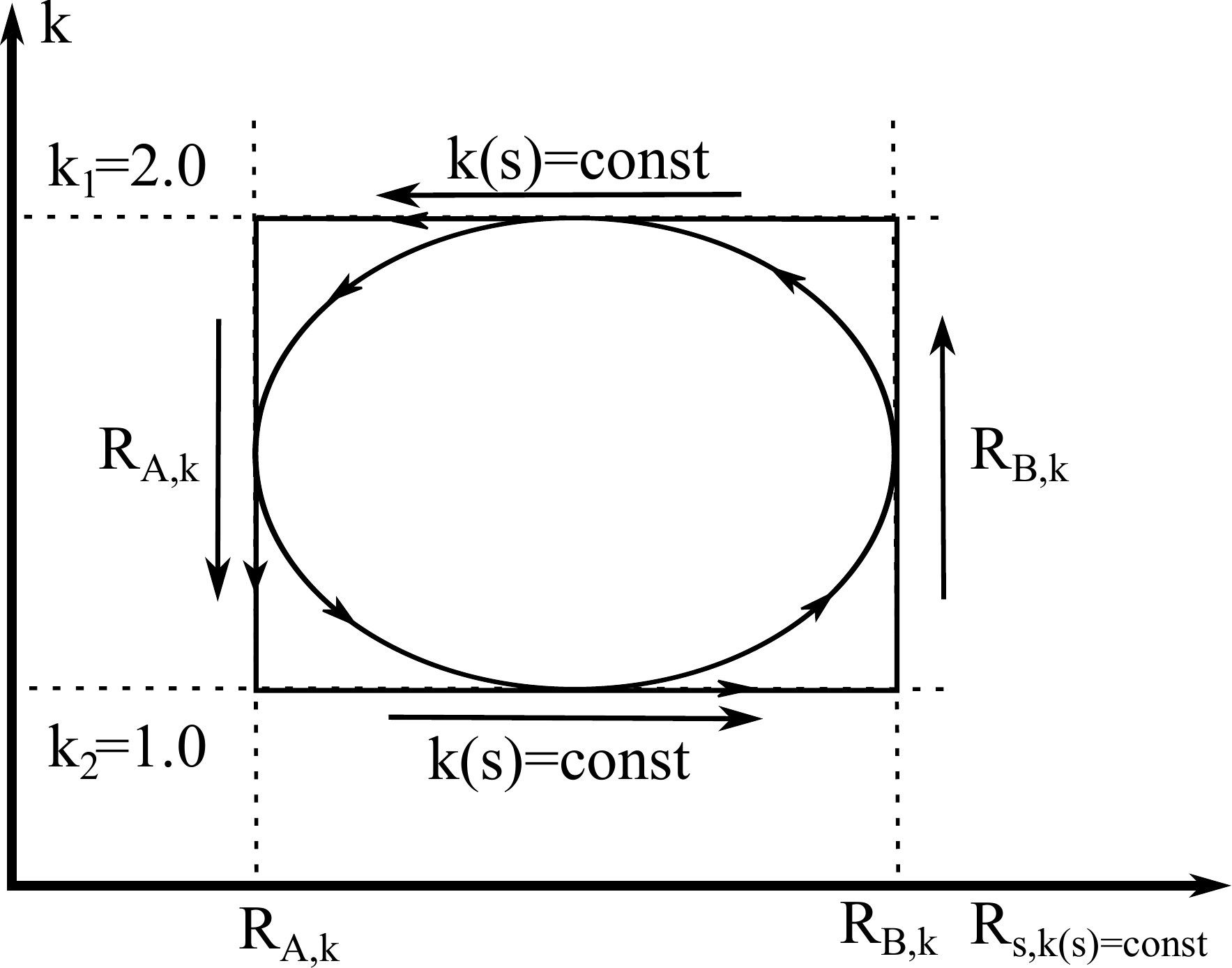}
  \caption{Representative contours of a closed loop in $k-R_k$ space
    including changes of regulator shape as well as RG scale. The
    rectangle contour separates solving the flow equations and
    switching the regulators. The ellipse contour however, allows for
    simultaneous changes in the scale and the regulator shape.}
\label{loopshape}
\end{figure}

\begin{center}
\begin{figure}[t!]
  \includegraphics[width=\columnwidth]{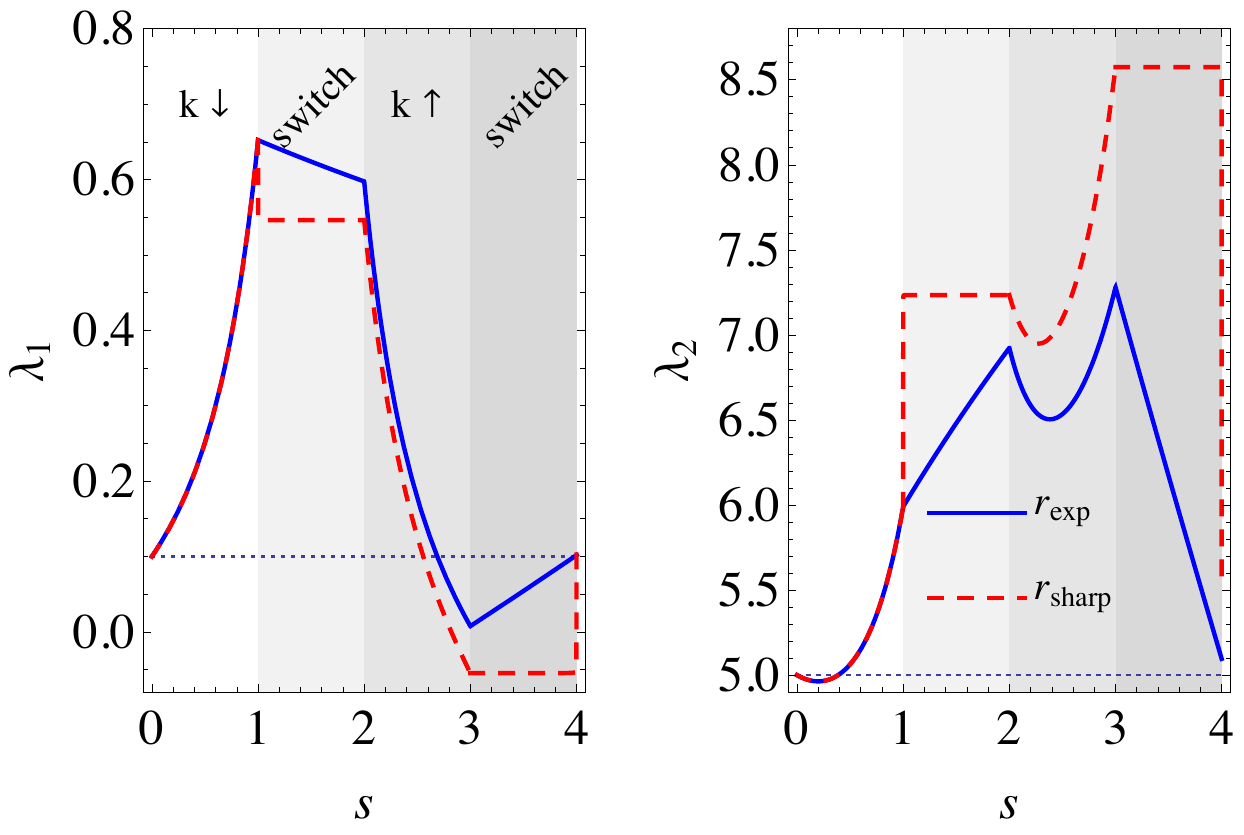}
  \caption{Flow of the couplings $\lambda_{1,2}$ along one closed-loop
    contour as defined in Fig.~\ref{loopshape}. The red dashed curve
    shows the switching from $r_{\text{flat}}$ to $r_{\text{sharp}}$
    and the blue solid curve the switch from $r_{\text{flat}}$ to
    $r_{\text{exp}}$. At the end of the closed-loop integration the
    values of the couplings $\lambda_{1,2}$ deviate slightly from
    their initial values.}
\label{1loop}
\end{figure}
\end{center}

\noindent In addition to the change of the regulator shape from
$R_{k}^{\text{A}}$ to $R_{k}^{\text{B}}$ at a fixed RG scale, we now
allow for a dependence of the RG scale $k$ on the loop variable $s$,
i.e., $k\rightarrow k(s)$. Then, we can perform integrations along
closed loops in theory space, cf. Fig.~\ref{fig:sflow}, and study the
violation of the integrability condition, \eq{eq:closedloop},
explicitly. Such a combined change of regulator and RG scale can be
incorporated in a linear interpolation between two scale dependent
regulators
\begin{align}
R_{s,k(s)}=a(s) R_{k(s)}^{\text{A}}+\big(1-a(s)\big)R_{k(s)}^{\text{B}}\,,
\end{align}
where $a(s)\in [0,1]$ parametrises the switching from one regulator
function to another.  In order to solve the flow equations along a
loop in $k-R_k-$ space we add to \eq{dulinsup} the terms which include
solving the flow equations in $k$ direction,
\begin{align}
\frac{d}{ds} u = J(u'+2\rho\, u'')+
\frac{\partial_s k(s)}{k(s)}\big(-d\,u+(d-2)\rho\, u'\big)\,.
\label{dulinsupB}
\end{align}
The $s$-derivative of the regulator in the threshold function
$J(\omega)$ defined in \eq{dulinsup} can be decomposed into two
contributions
\begin{align}
  \frac{d}{ds}r_s(y)=&\underbrace{\partial_s
    r_s(y)\textcolor{white}{\frac{|}{|}}}_{\text{change of shape}}\ -\
  \underbrace{2\,\frac{\partial_s k(s)}{k(s)}y\, \partial_y r_s(y)}_{
    \text{explicit change of scale}}\,,
\end{align}
where the first term keeps track of the change of the shape of the
regulator function $r_s(y)$, while the second term tracks the
change of the cutoff scale $k(s)$. Evidently this is just a convenient
splitting as the change of $k(s)$ can also be easily described by a
change of $r_s$. This is seemingly a trivial remark but it hints at
the fact that a change of the shape of the regulator may very well
imply a change of the physical cutoff scale. This discussion will be
detailed further in Section~\ref{sec:opt}. Explicitly, the involved
derivatives are given by
\begin{align}\nonumber 
\partial_s r_s(y)&=\big(r^{\text{A}}(y)-r^{\text{B}}(y)\big)\partial_s a(s)\,,\\[2ex] 
\partial_y r_s(y)&=a(s) \partial_y r^{\text{A}}(y)+(1-a(s))\partial_y
r^{\text{B}}(y)\,.
\end{align}
In the following, we again employ a linear superposition between two
regulators, e.g., $r_{\exp}$ and $r_{\text{sharp}}$, and solve the
flow of the scalar model in LPA to order $\phi^4$ along a closed loop.
Each closed loop in $k-R_k$ space along a rectangle contour then
consists of four steps, see Fig.~\ref{loopshape} for a representative
contour:
\begin{enumerate}
 	\item The flow equation is solved from $k_1$ to $k_2$. 
 	\item Switch the regulator continuously from $r^{\text{A}}$ to
          $r^{\text{B}}$.
 	\item Reverse the flow from $k_2$ to $k_1$.
 	\item Switch back from regulator $r^{\text{B}}$ to $r^{\text{A}}$.
\end{enumerate}
The flow of the couplings $\lambda_{1}$ and $\lambda_2$ along one such
closed loop is depicted in Fig.~\ref{1loop}.  For these calculations,
we again use the initial values $\lambda_1^{\text{(in)}}=0.1$ and
$\lambda_2^{\text{(in)}}=5.0$.  We switch from $r_{\text{flat}}$ to
$r_{\text{sharp}}$ (red dashed) or to $r_{\text{exp}}$ (blue solid),
respectively.  The change from $r_{\text{flat}}$ to $r_{\text{exp}}$
is smooth as both regulators are finite for all momenta. In contrast,
the change from $r_{\text{flat}}$ to $r_{\text{sharp}}$ shows a
discontinuous peak in the flow of $\lambda_{1,2}$: the transition from
the flat to the sharp regulator instantly lends an infinite infrared
mass to the propagator for momenta lower than the cutoff scale of the
sharp regulator. In either case the integration along one of our
chosen closed-loop contours shows slight deviations from the initial
values of $\lambda_1$ and $\lambda_2$, see Fig.~\ref{1loop}.

\begin{center}
\begin{figure}[t!]
\includegraphics[width=\columnwidth]{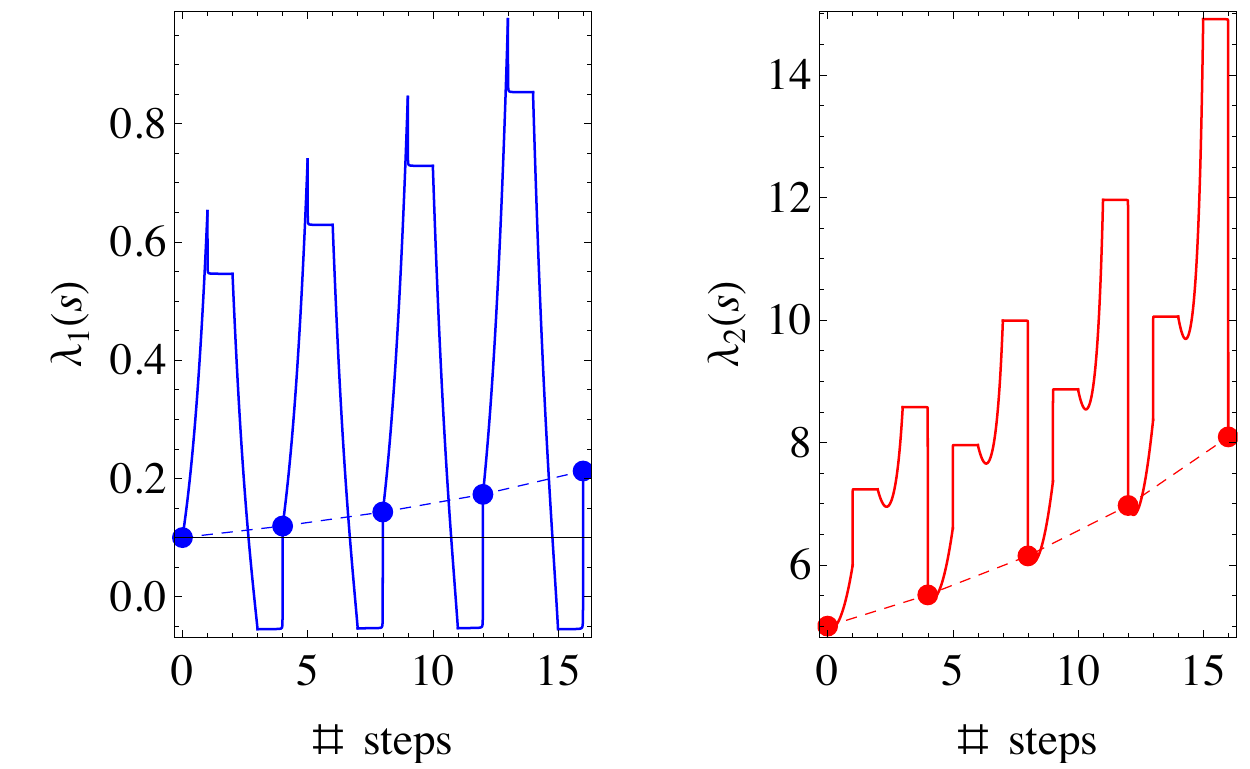}\\
\vspace{0.6cm}
\includegraphics[width=\columnwidth]{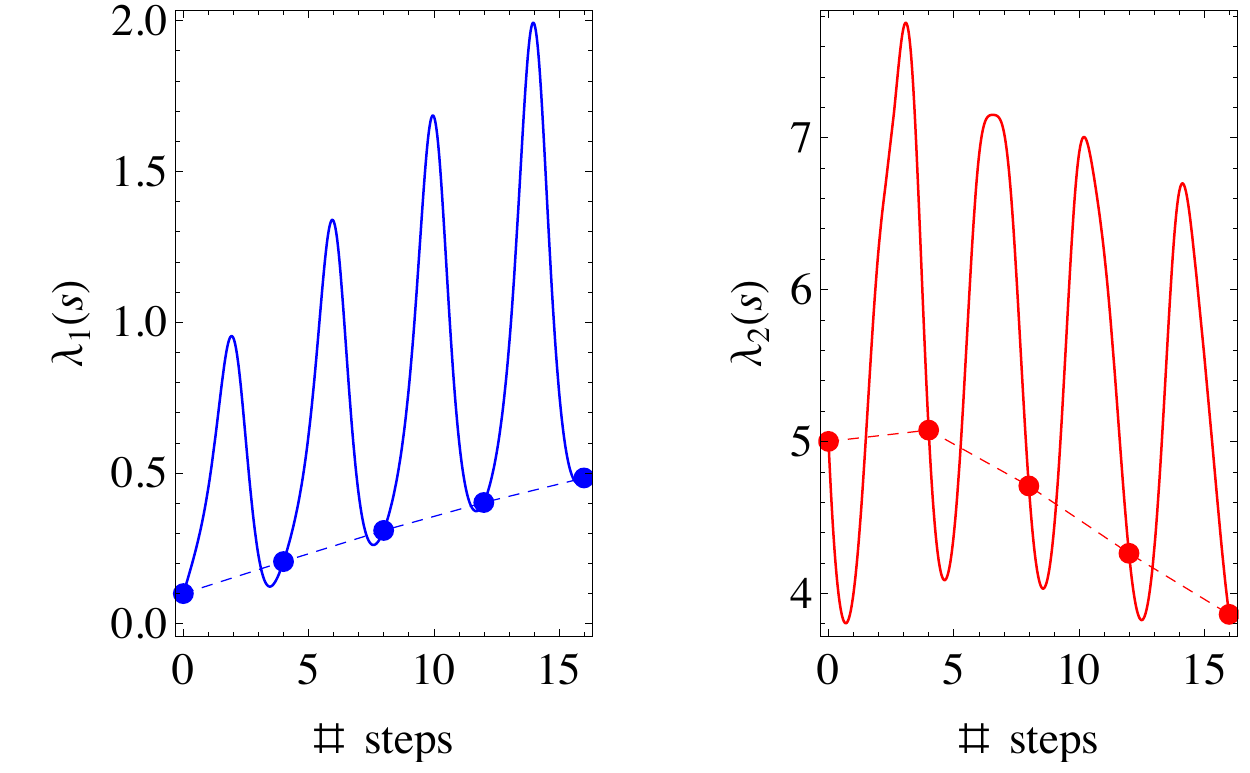}
\caption{Upper panels: Flow of the couplings $\lambda_{1,2}$ along
  four rectangular closed loops switching from $r_{\text{flat}}$ to
  $r_{\text{sharp}}$. Lower panels: Flow in $k-R_k$ space with
  $r_{\text{flat}}$ and $r_{\text{sharp}}$ along an ellipse contour
  (counter clockwise), with an interpolation starting with scale
  $k^{\text{(in)}}=2.0$ and
  $r=r_{\text{flat}}(y)+r_{\text{sharp}}(y/2)$,
  cf. App.~\ref{app:Exflow}.}
\label{4loops}
\end{figure}
\end{center}

The deviations from the initial values add up when the procedure of
integrating along a closed-loop contour in $k-R_k$ space is repeated.
This is shown in the upper panel of Fig.~\ref{4loops} for a
consecutive integration along four of the closed loops as defined in
Fig.~\ref{loopshape}.  In fact, after these four closed loops the
values of the coupling constants $\lambda_{1,2}$ strongly deviate from
their initial values.  For comparison, in Fig.~\ref{4loops}, we also
show an integration along an alternative closed-loop, defined by an
ellipse contour, cf. Fig.~\ref{loopshape}. This integration can be
performed in a completely analytical way for a transition from
$r_{\text{flat}}$ to $r_{\text{sharp}}$ as shown in 
App.~\ref{app:Exflow}.

Our study clearly demonstrates the violation of the integrability
condition, \eq{eq:closedloop}, for truncated renormalisation group
flows.  The severeness of this violation depends on the chosen
regulators, cf.~Fig.~\ref{1loop} and indicates the necessity of an
educated choice of the regularisation scheme in renormalisation group
investigations to establish and improve the reliability of physical
results.  The following section is dedicated to devising such an
educated choice in terms of an optimisation procedure.


\section{Optimisation}\label{sec:opt}
 
\noindent In order to obtain the best possible results from the
functional renormalisation group approach within a given truncation we
would like to single out the optimal regulator scheme for the
underlying systematic expansion. Here we follow the setup of
functional optimisation put forward in \cite{Pawlowski:2005xe}. The
discussion of systematic error estimates related to optimisation
requires a norm on the space of theories (at $k=0$) in order to
measure the severeness of the deviations. Here, we are not after a
formal definition but rather a practical choice of such a norm.

We illustrate complications with the definition of such a norm by
means of a simple example: we restrict ourselves to the local
potential approximation (LPA), or LPA${}^\prime$ where in the latter
we take into account constant wave function renormalisations $Z_k$.
Then, a seemingly natural choice is the cartesian norm on theory space
spanned by the constant vertices $\lambda_n =
\Gamma^{(n)}[\phi_{\text{\tiny{EoM}}}]$ evaluated, e.g., at the
equation of motion $\phi_{\text{\tiny{EoM}}}$.  However, this falls
short of the task as it weights a deviation in higher correlations or
vertices $\lambda_n$ in the same way as that of the lower ones,
despite the fact that the lower ones are typically more
important. Additionally, the $\Gamma^{(n)}$ are neither
renormalisation group invariant nor do they scale identically, see
\eq{eq:RGGn}.

If we extend the above setting to a general vertex expansion scheme,
the coordinates in theory space are related to
$\Gamma^{(n)}[\phi_{\text{\tiny{EoM}}}](p_1,...,p_n)$.  These
quantities are operators and the definition of the related $n$th axis
of the coordinates system requires a suitably chosen operator norm,
for a more detailed discussion see \cite{Pawlowski:2015mia}. Even
though this general case can be set-up, for most practical purposes it
is sufficient to rely on a simple definition of a norm adapted to the
approximation at hand.  

Let us assume that we found a norm that allows to define the length
$L[\CC]$ of a given flow along a trajectory $\CC$ in theory space
parameterised with $s\in[0,1]$, flowing from some regulator $R_{s=0}$
to $R_{s=1}$.  For example, we can consider the global flow with a
given regulator from $k=\Lambda$ to $k=0$, i.e., the flow trajectory
does not necessarily have to be a closed loop.  The discussions in the
previous section suggest that, in a given approximation, we should try
to minimise this length in order to minimise the systematic error.
Accordingly, we have to compare the lengths of different trajectories
$L[\CC]$.  This heuristic argument can be made more precise,
\cite{Pawlowski:2015mia}: without approximation the final effective
action $\Gamma[\phi,R=0]$ does not depend on the trajectory, in other
words
\begin{align}
\0{\delta}{\delta R_s(p)}\int_0^1 ds\, \partial_s \Gamma[\phi,R] =0\,.
\end{align}
Note however, that this discussion bears an
intricacy, as it implies the comparison of the length of flow
trajectories of physically equivalent effective actions
$\Gamma[\phi,R^{\text{A}}]$ and $\Gamma[\phi,R^{\text{B}}]$ towards
$\Gamma[\phi,0]$.
Therefore, we should compare
trajectories that always start at physically equivalent effective
actions at a large physical cutoff scale.


\subsection{Physical Cutoff Scale}

\noindent The cutoff parameter $k$ is usually identified with the physical
cutoff scale, but such an identification falls short in the general
case. To understand this, let us re-evaluate the example of the
flows with $r^{\text{A}}$ and $r^{\text{B}}$ leading to the circular
flow \eq{eq:intcond}.  In the spirit of the discussion above it seems
to be natural to compare the two flows from $k=\Lambda$ to $k=0$ with
the regulators $r_{s=0}=r^{\text{A}}$ and $r_{s=1}=r^{\text{B}}$,
respectively, while the $s$-flow in this example simply switches the
regulator at a fixed scale $k=\Lambda$.  This picture fails trivially
for
\begin{align}\label{eq:r2} 
r_1(x)=  r_0(x/\lambda)\,, \quad {\rm with} \quad 
R_{1,k}=R_{0, \lambda k}\,, 
\end{align}
where the change of regulators simply amounts to changing the scale.
As trivial as this example is, it hightlights a key question:\\

\noindent {\it What is
the physical cutoff scale for a given regulator?}\\

\noindent In Ref.~\cite{Pawlowski:2005xe} it has been argued that
within practical applications it is suggestive to use the physical gap
of the theory as the practical definition of the physical infrared or
cutoff scale.  Strictly speaking this asks for the evaluation of the
poles and cuts of the theory in a real time formulation. For the
scalar and Yukawa-type theories explicitly discussed in the present
work it has been shown in \cite{Helmboldt:2014iya} that the real-time
pole masses and the imaginary time curvature masses are very similar
in advanced approximations. For the present purpose these subtleties
are not relevant and we define the inverse gap as the
maximum of the imaginary time propagator,
\begin{align}\label{eq:kphys}
  \0{1}{k_{\text{\tiny{phys}}}^2} = \max_{p,\phi}\, G[\phi,R]\,.
\end{align}
For the sake of simplicity we have restricted ourselves to constant
backgrounds $\phi$. In the general case, \eq{eq:kphys} picks out the
maximal spectral value of the propagator $G$~\cite{Pawlowski:2005xe}.
Note also that in theories with several fields one has to monitor the
gaps of all the fields involved. In the present work this is important
within the example theories studied: the relativistic $O(M)\oplus
O(N)$ models as a simple model theory, as well as a non-relativisitc
Yukawa model for impurities in a Fermi gas. A further intricacy
originates in different dispersion relations of the fields involved
such as relativistic scalar field with $p^2$ and fermionic fields
with $\ip$. Then, the relative physical cutoff scale may involve a
nontrivial factor in comparison to the gap. The latter subtlety will
be discussed elsewhere.

\begin{figure}[t!]
  \includegraphics[width=0.83\columnwidth]{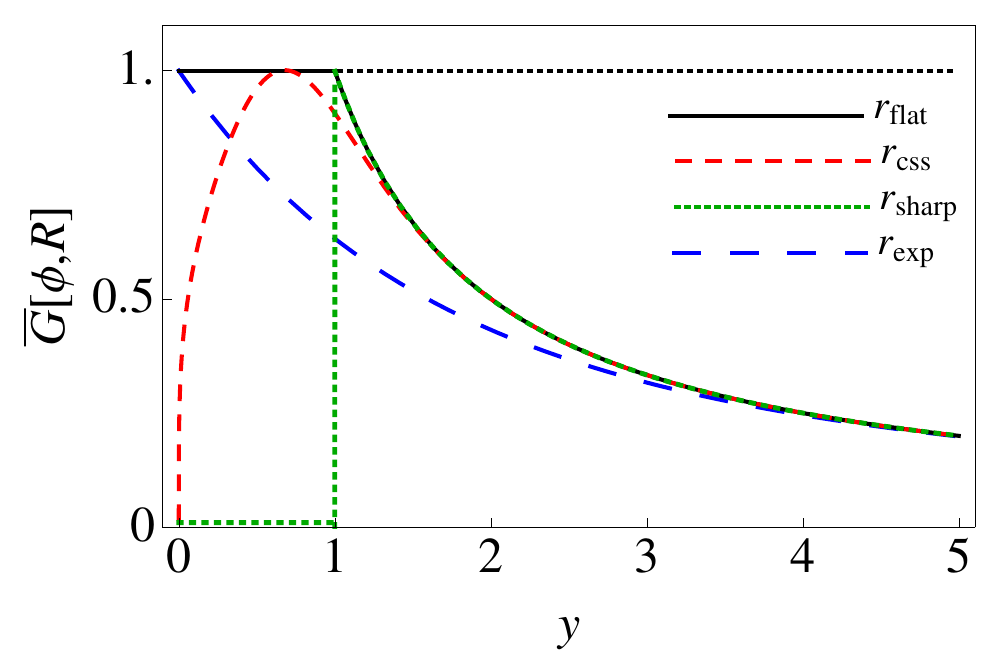}\\
  \caption{Comparison of different normalized and regularized
    propagators $y(1+r_i(y))^{-1}$ with shape functions $r_i(y)$ as
    listed in Tab.~\ref{tregs}. Here, $r_{\text{css}}$ is plotted with
    $b=1.25,c=3$ and $h=0.12$.}
\label{fig:propnorm}
\end{figure}

Note also, that in \eq{eq:kphys} a fixed identical RG scheme for all
regulators is required, as the propagator is not invariant under RG
transformations $\partial_s G[\phi,R] = - 2 \gamma_\phi G[\phi,R]$,
cf.~App.~\ref{app:rgtrafo}. Such a fixed RG scheme can be defined by first
selecting one specific flow from $k_{\text{\tiny{phys}}}=\Lambda$ to
$k_{\text{\tiny{phys}}}(k=0)$, the latter being the physical gap of
the theory at $k=0$. Then regulator changing flows such as defined in
\eq{eq:flow12} at the fixed physical UV scale
$k_{\text{\tiny{phys}}}=\Lambda$ lead to initial conditions within the
same RG scheme defined at $k=0$.  This leads to closed flows without
taking into account a further RG transformation at $k=0$. Hence,
\eq{eq:kphys} implies that the normalised dimensionless propagator
satisfies a renormalisation group invariant bound,
\begin{align}\label{eq:<1}
  \bar G[\phi,R]\leq 1\, \forall\, p,\phi\,,\ \ \text{with}\quad \bar
  G[\phi,R]= k_{\text{\tiny{phys}}}^2 G[\phi,R]\,.
\end{align}
In summary, we call theories in the presence of a regulator physically
equivalent, if the gaps $k_{\text{\tiny{phys}}}$ of all fields agree.
In \fig{fig:propnorm} we present some examples for this criterion for
classical propagators in a theory with
$V''(\phi_{\text{\tiny{min}}})=0$.  These examples are relevant for
the LPA approximation which we predominantly use in the present work.
%


\subsection{Optimisation and Length of a RG Trajectory}\label{sec:RGlength}

\noindent Now we are in the position to define the length of a flow trajectory
$\CC$.
Keeping in mind the discussion of the coordinate system in theory  
space at the beginning of this Section, we reduce the task by
using the effective action itself, evaluated on fields close to the
solution of the quantum equations of motion $\phi_{\text{\tiny{min}}}$
with
\begin{align}\label{eq:QEoM}
  \left.\0{\delta\Gamma[\phi,R]}{\delta
      \phi}\right|_{\phi=\phi_{\text{\tiny{min}}}}=0\,.
\end{align}
The value of the effective action has no physics interpretation and
depends on the renormalisation procedure, i.e., the chosen regulator
and initial condition.  Therefore, we resort to the second derivative
$\Gamma^{(2)}[\phi,R]$ rather than to $\Gamma[\phi,R]$ itself.
Indeed, the natural choice is the connected two-point correlation
function or rather the normalised dimensionless two-point function
$\bar G[\phi,R]= k_{\text{\tiny{phys}}}^2 \langle
\phi(p)\phi(-p)\rangle_c$, cf. \eq{eq:<1}. Here, the subscript $c$
refers to the connected part.  This is motivated by the fact that the
master equation \eq{eq:flow} only depends on the propagator, as do the
operator representations for the total $t$- and $s$-derivatives,
\eq{eq:repdt} and \eq{eq:repds}.

Measuring the length of the flow of the dimensionless
propagator $\bar G[\phi,R]$ requires a coordinate system in theory
space where the axes are, e.g., expansion coefficients of the
propagator, $\bar G^{(n)}$ or the spectral values of the propagators
linked to an expansion in the eigenbasis of $\bar G$
\begin{align} 
  \bar \lambda_{\text{\tiny{max}}}= 1\,,\quad {\rm for}\quad \bar
  \lambda\in {\rm spec} \,\bar G[\phi, R(k_{\text{\tiny{phys}}})]\,.
\end{align}
To summarise, the procedure is choosing an operator norm $\|.\|$ for
$\bar G$ as well as for $\partial_s \bar G$.  Then, we define the
length of a flow trajectory $\CC$ with that of the length of
$\|\partial_s \bar G\|$.

Before we come to integrated flows, let us evaluate the consequences
of the discussion above.  Firstly, we assert that monotonous flows are
shorter than non-monotonous ones.  Assuming already a restriction to
monotonous flows for $G$ and hence $\bar G$, we find a simple
optimisation criterion in terms of the dimensionless propagator: $\bar
G$ is bounded from above by unity, see \eq{eq:<1}.  Moreover, for
optimal regulators the propagator is already as close as possible to
this bound due to its monotonous dependence on $t$.  This leaves us
with
\begin{align}\nonumber 
 &\|\bar p^2 (\bar G[\phi, R_{\text{\tiny{opt}}}]-\bar G[\phi,0])\|\\[2ex] 
&\quad\quad = 
  \min_{R\in R( k_{\text{\tiny{phys}}})}\|\bar p^2(\bar G[\phi,
  R]-\bar G[\phi,0])\|\,,
\label{eq:minbarG}
\end{align}
where $\bar p^2 = p^2/k_{\text{\tiny{phys}}}^2$. The prefactor $\bar
p^2$ has been introduced for convenience to easily accommodate also for
massless modes at vanishing cutoff scale. The criterion \eq{eq:minbarG}
has been derived in \cite{Pawlowski:2005xe}, where it also has been
shown that for optimised regulators local integrability is restored.

With \Eq{eq:minbarG}, for a given background $\phi$, we have reduced
theory space to a one-dimensional subspace with a simple cartesian
norm. Still, the space of regulators is infinite-dimensional and the
length of a given flow curve parametrised by $s$ is related to the
size of the flow operator equations~\eq{eq:repdt} and \eq{eq:repds}
for $t$-flows or $s$-flows, respectively.  The flow operators involve
second-order $\phi$~derivatives as well as kernel of the flow
operator,
\begin{align}\label{eq:t}
\flow[\phi,R] = G[\phi,R]\,\partial_s R\,G[\phi,R]\,.
\end{align}
The $\phi$~derivatives act on the complete set of observables and
their action is general. Therefore, we simply have to integrate the
size of $\flow[\phi,R]$ along the flow for
computing a relevant length.
For constant backgrounds $\phi$ we integrate over all
spectral values of the operator
\begin{align}\label{eq:specnorm}
\| \flow[\phi,R]\|= \int_0^\infty d p^2 \,|\flow [\phi,R]|\,,   
\end{align}
giving a dimensionless quantity.  This spectral definition can be
extended to general backgrounds.  Moreover, it can be extended to more
general norms that, e.g., take into account the importance of smooth
regulators for the derivative expansion \cite{Pawlowski:2005xe}.  The
norm in \eq{eq:specnorm} diverges for $\flow[\phi,R]$ showing a
infrared singularity with more than $1/p^2$.  This can be amended with
additional powers of $\bar p^2$.

In summary this leads us to the final expression for the length of a
trajectory at a given value of $V''(\phi)$,
\begin{align}\label{eq:flowlength} 
L[V'',R]= \int_0^1 ds\,\sqrt{1+\| \flow[\phi(V''),R]\|^2}\,, 
\end{align}
where $\phi(V'')$ is chosen such that $V''(\phi(\omega))$ is
fixed.
Then, \eq{eq:flowlength} is the length of the trajectory
for $G$,
\begin{align}\label{eq:Glength} 
 \int_0^1 \sqrt{d s^2 +\| d G[\phi(V''),R]\|^2}=L[V'',R]\,,
\end{align}
where we have used that $\partial_s V''\equiv 0$. With \eq{eq:Glength}
the optimisation criterion \eq{eq:minbarG} now can be recast into
\begin{align}\label{eq:Optlength} 
  L[V'',R_{\text{\tiny{opt}}}]= \min_{R\in R(
    \Lambda_{\text{\tiny{phys}}})}L[V'',R]\,,
\end{align}
where $\Lambda_{\text{\tiny{phys}}}$ indicates that all flows start at
the same physical scale. Note that identical physical UV scales are
typically easily identified. Hence, for global flows from the
ultraviolet to the infrared we have trajectories with $\Gamma_\Lambda
= \Gamma[\phi,R_\Lambda]$ with $R(s=0)=R_\Lambda$, and
$\Gamma_{k=0}=\Gamma[\phi,0]$ with $R(s=1)\equiv 0$.  The optimal
regulator should minimise the length of the flow $L[V'',R]$ for all
$\phi$.  A comparison of the length for different regulators will be
presented in the following Section~\ref{sec:pracopt}.

We close with the remark that both criteria, \eq{eq:minbarG} and
\eq{eq:Optlength}, implement the functional optimisation criterion
from \cite{Pawlowski:2005xe}, and hence are identical. In practical
applications the one or the other may be more easily accessible.


\subsection{Practical Optimisation}\label{sec:pracopt}

\noindent Let us exemplify the above construction at the
example of the LPA approximation for one real scalar field.
Its propagator for a given gap
$k_{\text{\tiny{phys}}}$ reads
\begin{align}\label{eq:propLPA} 
  G[\phi_{\text{\tiny{min}}},R]= \0{1}{p^2 +\omega_{\text{\tiny{min}}}+
    R}\,.
\end{align}
where it is understood that the cutoff scales in the regulator $R$ is
adjusted such that the maximum of the propagator is
$1/k_{\text{\tiny{phys}}}^2$, and $\omega_{\text{\tiny{min}}}=V''(\phi_{\text{\tiny{min}}})$ stands for the
curvature at the minimum of the effective potential.
Now, we use that an optimised regulator minimises
infinitesimal flows as well as the rest of the flow towards $k=0$.
This statement holds for correlation functions and, in particular, for the
propagator entailing that the difference between the optimal
propagator for a given physical cutoff
scale $k_{\text{\tiny{phys}}}$ and the propagator at $k=0$  is minimal.

Let us assume for the
moment that $\omega_{\text{\tiny{min}}}$ is 
already at the value it acquires at $k=0$.
Then, we are left with the condition to minimise
\begin{align}\label{eq:opt}
  &|G[\phi_{\text{\tiny{min}}},R] -G[\phi_{\text{\tiny{min}}},0] |  \nonumber\\
  &\quad\quad\quad\quad=
  \left| \0{R}{(p^2 +\omega_{\text{\tiny{min}}}+ R)(p^2
      +\omega_{\text{\tiny{min}}})}\right|\,,
\end{align}
for all momenta with the constraint \eq{eq:kphys}.
We now make a further simplification and set $\omega_{\text{\tiny{min}}}=0$. Then, we
are left with minimising
\begin{align}\label{eq:opt1}
  \left|\0{r}{(p^2 + R)}\right|\,,
\end{align}
for all momenta.
For momenta $p^2 \geq k_{\text{\tiny{phys}}}^2$ we
immediately arrive at $ r_{\text{\tiny{opt}}}=0$. 
For $p^2 < k_{\text{\tiny{phys}}}^2$ the regulator has to be positive in order to
account for the gap condition \eq{eq:kphys}.
If this condition is saturated, \eq{eq:opt} is minimised, 
leading to $p^2 + p^2
r_{\text{\tiny{opt}}}= k_{\text{\tiny{phys}}}^2$, and hence
$r_{\text{\tiny{opt}}} = k^2_{\text{\tiny{phys}}} /p^2 -1 $ for the
momenta $p^2 < k_{\text{\tiny{phys}}}^2$.
In combination with the
vanishing for $p^2 \geq k_{\text{\tiny{phys}}}^2$ this leads to the
unique optimised regulator in LPA,
\begin{align}\label{eq:ropt}
  r_{\text{\tiny{opt}}}= \left(\0{k^2_{\text{\tiny{phys}}} }{p^2}
    -1\right)\theta(k^2_{\text{\tiny{phys}}} -p^2)\,,
\end{align}
the flat or Litim
regulator~\cite{Litim:2000ci,Litim:2001up,Litim:2001fd}.  Note that
there it has been introduced as one of a set of optimised regulators,
being distinct by its analytic properties.  It has been singled out as
the unique solution of the functional optimisation in
Ref.~\cite{Pawlowski:2005xe}.  Indeed, the critical exponents in
$O(N)$ theories truncated in a local potential approximation with this
regulator are closest to the physical ones.  The above simplified
derivation can be upgraded to also take into account a given fixed
$\omega_{\text{\tiny{min}}}$. Again this leads to \eq{eq:ropt} with
$k_{\text{\tiny{phys}}}^2\to k^2$ where $k^2$ runs from
$k_{\text{\tiny{phys}}}^2-\omega_{\text{\tiny{min}}}$ to zero.

\begin{figure}[t!]
  \centering
  \includegraphics[width=0.95\columnwidth]{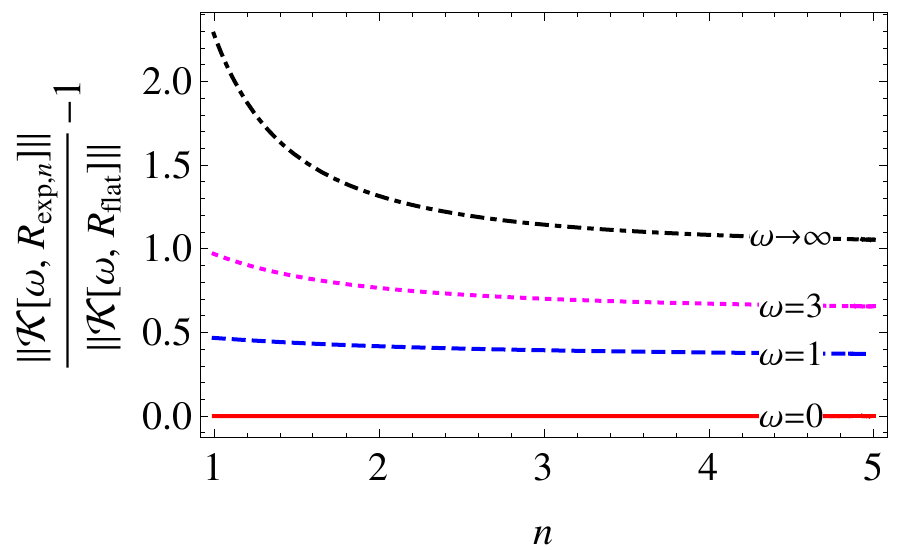}
  \caption{Deviation of the ratio of
    $\|\flow[\omega,R_{\text{\tiny{exp}},n}]\|/\|\flow[\omega,R_{\text{\tiny{flat}}}]\|-1$,
    as a function of the power $n$ in the exponential, for
    $\omega\in[0,\infty)$. For $\omega\to \infty$ the bound
    saturates. For $\omega=0$ the norm is a total derivative and is
    independent of the choice of regulator, hence the ratio is unity:
    $\omega=0$: red straight line, $\omega=1$: blue dashed line,
    $\omega=3$: magenta dotted line, $\omega=\infty$: black
    dashed-dotted line. }
\label{fig:tnorm}
\end{figure}
\begin{figure}[t!]
  \centering
  \includegraphics[width=0.95\columnwidth]{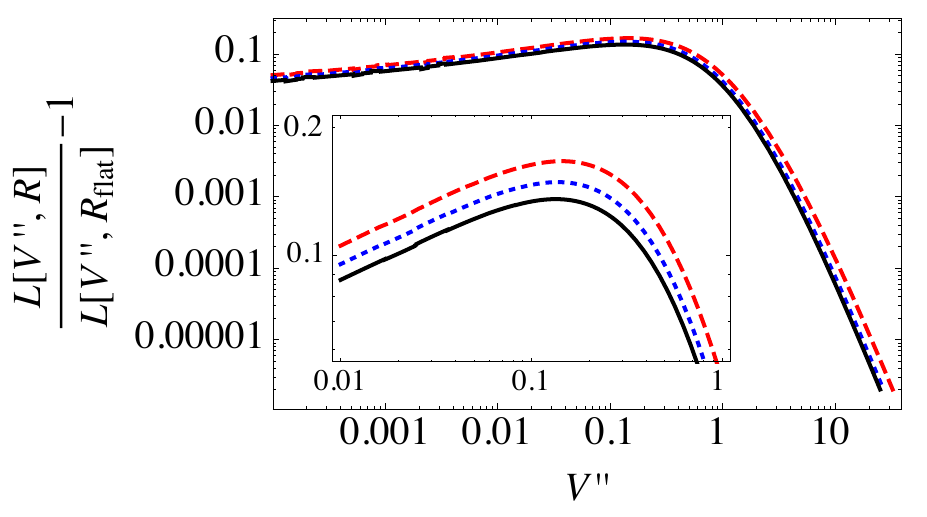}
  \caption{Length $L[V'',R]/L[V'',R_{\text{\tiny{flat}}}]-1$ for
    different exponential regulators in comparison to the flat one:
    $n=1$: red dashed line, $n=2$: blue doted line, $n=4$: black
    straight line. The length is minimised for the flat regulator.}
\label{fig:flowlength}
\end{figure}

Let us now also compare the lengths of the trajectories as defined in
Sec.~\ref{sec:RGlength}.  The definition was adjusted such that it
does not require the knowledge about $\Gamma^{(2)}[\phi,R]$ along the
flow leading to simple practical computations.  A more elaborate
version of the present case is straightforwardly implemented by
relaxing $\partial_s V''\neq 0$.

In \fig{fig:tnorm} we first compare the norms, \eq{eq:specnorm}, of
the flow operator for different values of $\omega=V''/k^2$.  We show
the deviation of the ratios
$\|\flow[\omega,R_{\text{\tiny{exp}},n}\|/\|\flow[\omega,R_{\text{\tiny{flat}}}]\|-1\geq
0$ from one for all values of $\omega$.  Here,
$R_{\text{\tiny{exp}},n}$ are the exponential regulators with the
corresponding shape function
\begin{align}\label{eq:expn}
r_{\text{\tiny{exp}},n}(y) = \0{y^{n-1}}{e^{y^n}-1}\,,
\end{align}
which is a specific subclass of the exponential interpolating
regulator in Tab.~\ref{tregs} with $a=1, b=0$.  The deviation is
always bigger than zero, which singles out the flat regulator as the
optimal one in LPA, see \fig{fig:tnorm}.  Note in this context that
\fig{fig:tnorm} gives us the full information of the relative size of
the integrands in the length of the flow in \eq{eq:flowlength}: for a
given $V''>0$ the related $\omega$ diverges with
$1/k_{\text{\tiny{phys}}}^2$. This is easily confirmed with the
explicit computation of the length, summarised in
\fig{fig:flowlength}, where the global length is shown for given
$V''$. One can observe that the flat regulator minimises the flow
length $L[V'',R]$ which supports the optimisation criterion developed
in Sec. \ref{sec:RGlength}.

Note also that the norms are defined such that the information about
the physical scales
$k^2_{\text{\tiny{phys}}}(s=1)/k^2_{\text{\tiny{phys}}}(s=0)$ is only
encoded in $\omega(s=1)/\omega(s=0)$. Hence, for large $V''$ in
comparison to the physical cutoff scales the difference between the
different flows is large. However, in this regime the absolute size of
the flow is small.


\section{Critical Behavior of Multi-Field Models}\label{sec:multifield}

\noindent Many interesting systems include a collection of different
field degrees of freedom.  In this situation the choice of suitable
combinations of regulators is not straightforward and we have already
mentioned the relativistic Yukawa models with structurally different
dispersion for scalars and fermions.  Here we study this case within a
simple situation, a bosonic $O(M)\oplus O(N)$ model.  We show that
the choice of relative cutoff scales generally has a crucial impact on
the obtained results for the critical physics: The $O(M)\oplus O(N)$
model has two competing order parameter fields and the competing
order makes it particularly sensitive to small effective changes of
the relevant parameters. The model is studied in the lowest order of
the derivative expansion, in LPA. It is well known that such a
truncation already captures well the critical physics of scalar models
despite the lack of non-trivial momentum dependences of
propagators and vertices. The latter encode the anomalous dimensions
of the system which are quantitatively small, here, and
hence can be neglected. 

However, the momentum dependences are also important for taking into
account the momentum transfer present in the diagrams on the right
hand side of the flow equation. For identical physical cutoff scales
this momentum transfer is minimised. In turn, for shifted relative
physical cutoff scales of different fields the diagrams have a sizeable
momentum transfer. In such a case, physics that is well incorporated
in the LPA with identical physical cutoff scales, is lost if the
difference between the physical cutoff scales grows large. If one goes
beyond LPA within systematic momentum-dependent approximation schemes
this relative cutoff scale dependence eventually disappears. The
discussion also emphasises the necessity of identical physical cutoff
scales within a given approximation in the sense of an optimisation of
approximation schemes.

In the present section we highlight the physics changes that are
triggered by the change of relative cutoff scales in LPA. As discussed
above, due to the missing momentum dependences of LPA, different
relative cutoff scales effectively lead to different actions at $k=0$,
see also \Fig{fig:spread}. In LPA, the bosonic, $d$-dimensional
$O(M)\oplus O(N)$ model has the following effective action,
\cite{Eichhorn:2013zza,Eichhorn:2014asa,Eichhorn:2015woa},
\begin{align}
  \Gamma_k=\int_x \left[ \frac{1}{2} (\partial_\mu\phi)^2 +\frac{1}{2}
    (\partial_\mu\chi)^2 +U_k(\phi,\chi) \right] \label{eq14}\,,
\end{align}
where $\phi$ and $\chi$ are $N$- and $M$-component fields,
respectively.  The effective potential 
$U_k(\bar\rho_\phi,\bar\rho_\chi)$ only depends on the field
invariants $\bar\rho_\phi=\phi^2/2$ and $\bar\rho_\chi=\chi^2/2$.  The
scale-dependent dimensionless effective potential is given by
\begin{align}\label{eq:umult}
  u=u(\rho_\phi,\rho_\chi)=k^{-d}U_k(\bar\rho_\phi,\bar\rho_\chi)\,\quad
  {\rm with}\quad \rho_i=k^{2-d}\bar\rho_i\,,
\end{align} 
and $i\in \{\phi,\chi\}$.  We further introduce the shape functions
$r_\phi$ and $r_\chi$ to regularise the $\phi$ and $\chi$ field modes,
respectively. The flow equation for the dimensionless effective
potential \eq{eq:umult} reads
\begin{align}
  \partial_t{u}=&-d\,u+(d-2)\rho_{\phi} u^{(1,0)}+(d-2)
  \rho_{\chi} u^{(0,1)}\nonumber\\[2ex]
  &+I_{R,\phi}(\omega_\phi,\omega_\chi,\omega_{\phi \chi})
  +(N-1)I_{G,\phi}(u^{(1,0)}) \nonumber \\[2ex]
  &+I_{R,\chi}(\omega_\phi,\omega_\chi,\omega_{\phi
    \chi})+(M-1)I_{G,\chi}(u^{(0,1)})\,,
 \label{eq29}
\end{align}
where we have introduced suitable threshold functions
$I_{i,j}(x,y,z),\ i \in\{R,G\},\ j \in\{\phi,\chi\}$ to separate the
loop integration over the radial and Goldstone modes for the two
fields.  The explicit expressions for these threshold functions are
listed in App.~\ref{app:thresh}.  The arguments of the threshold
functions are given by $\omega_\phi=u^{(1,0)}+2\rho_\phi u^{(2,0)}$,
$\omega_\chi=u^{(0,1)}+2\rho_\chi u^{(0,2)}$ and
$\omega_{\phi\chi}=4\rho_\phi\rho_\chi \big(u^{(1,1)}\big)^2$.  For
calculations, we expand the effective potential about the flowing
minimum $(\kappa_\phi,\kappa_\chi)$, to wit 
\begin{align}
  u(\rho_\phi,\rho_\chi) = \sum_{1 \leq i+j \leq \textrm{ord} }
  \frac{\lambda_{ij}}{i!j!}(\rho_\phi-\kappa_{\phi})^{i}(\rho_\chi-\kappa_{\chi})^j\,.
\end{align}
In this truncation we follow the flow of the couplings $\kappa_i$ and
$\lambda_{ij}$ which are given in
App.~\ref{app:twofieldcouplings}. These $O(N)\oplus O(M)$ models
posses a rich variety of fixed points exhibiting different types of
multi-critical behaviour relevant to a number of physical systems
\cite{PhysRevB.13.412,PhysRevB.8.4270,PhysRevLett.33.813,PhysRevB.67.054505}. For
our further investigations, we list the properties of the three most
important fixed points:\\[-2ex]

(i) The \emph{decoupled fixed point} (DFP) is characterised by a
decomposition into two disjoint $O(N)$ and $O(M)$ models where all
mixed interactions vanish, e.g., $\lambda_{11}=0$. It inherits all of
the properties of the Wilson-Fisher (WF) fixed points of the separate
sub-sectors.\\[-2ex]

(ii) The \emph{isotropic fixed point} (IFP) features a
symmetry enhancement where at each order in the fields the couplings
are degenerate, e.g.,
$\lambda_{20}=\lambda_{02}=\lambda_{11}$. Therefore, the fixed point
coordinates agree with the ones from an $O(N+M)$ symmetric model and
it inherits all of its critical exponents.\\[-2ex]

(iii) The \emph{biconical fixed point} (BFP) is a non-trivial fixed
point with interactions in both sectors that does not provide a
symmetry enhancement. This fact makes it interesting for our further
analysis, because it can be easily distinguished by means of
the critical exponents of a single field model.\\[-2ex]

We have listed values for the largest critical exponent $y_1=1/\nu_1$
for the models and fixed points which are important to this work in
Tab.~\ref{relcCritExp}, showing results for different levels of the
truncation.  Generally, we sort the critical exponents according to
the definition $y_1>y_2>y_3>...\,$.

\begin{table}[t!]
\[
\begin{array}{c|c|c|c|c|c}
\hline\hline
\text{Fixed point} &\rho^2&\rho^3&\rho^4&\rho^5&\rho^6\\
\hline\hline
\text{BFP in } \ \text{O}(2)\oplus\text{O}(2)& 1.78&1.61& 1.78&1.77&1.75\\
\hline
\text{BFP in }  \ \text{O}(1)\oplus\text{O}(2)&
2.16&1.42 & 1.65 &1.63&1.62\\
\hline
\text{WF in O}(2)& 1.61 & 1.32&1.40&1.42&1.41  \\
\hline
\text{WF in O}(1)& 2.00 &1.37& 1.54&1.55& 1.54\\
\hline\hline
\end{array}
\]
\caption{List of values for the largest critical exponent $y_1$ for 
  selected fixed points in LPA to ascending order in the truncation.}\label{relcCritExp}
\end{table}
\begin{figure*}[t!]
  \centering
  \includegraphics[width=0.68\columnwidth]{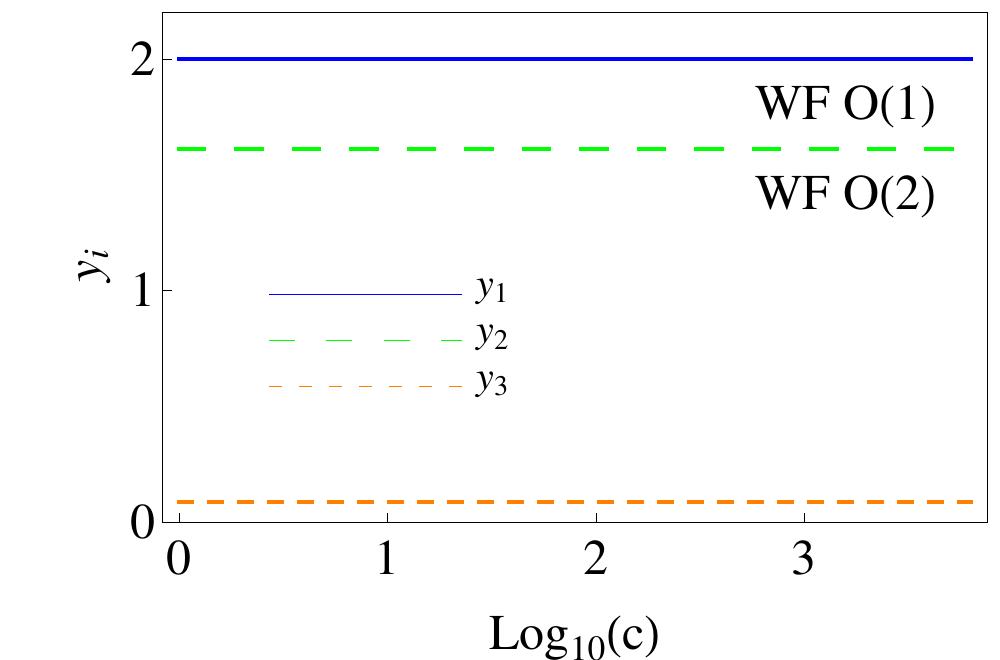}
  \includegraphics[width=0.68\columnwidth]{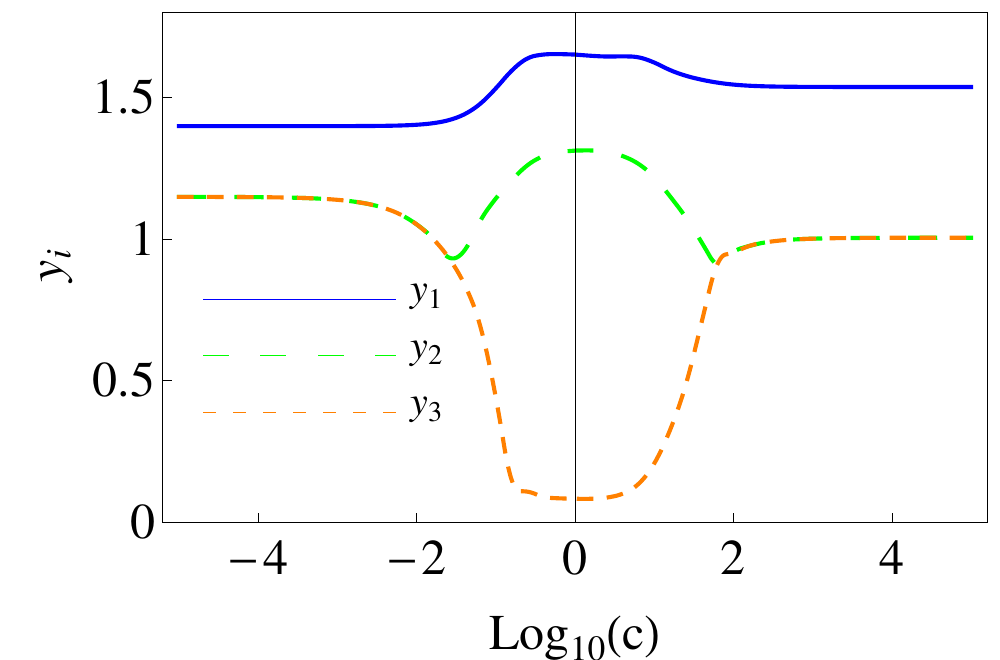}
  \includegraphics[width=0.68\columnwidth]{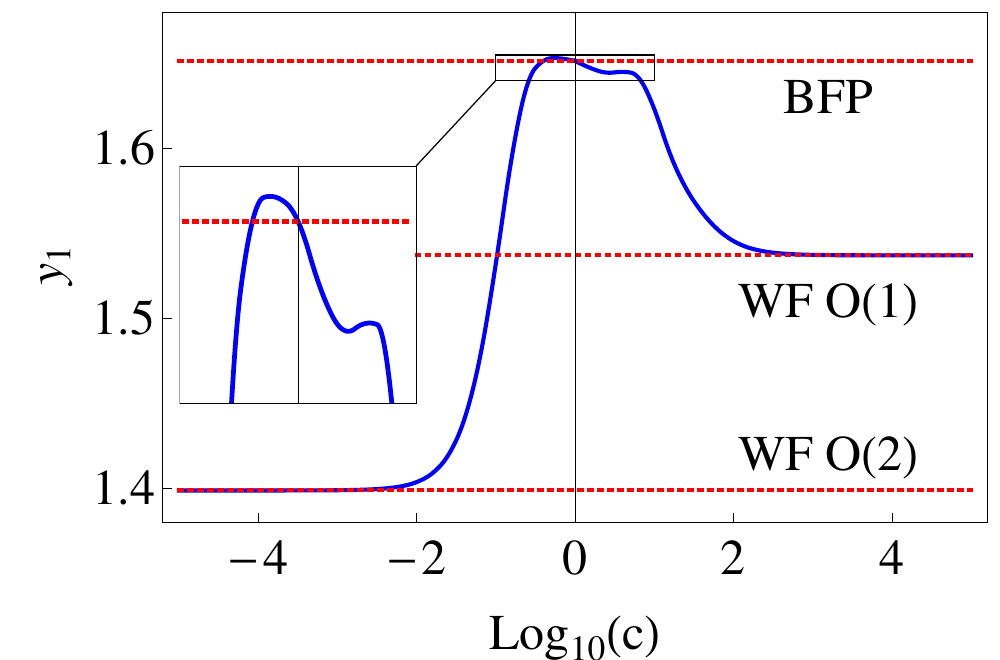}
  \caption{Left panel: DFP critical exponents of the
    $\text{O}(2)\oplus\text{O}(1)$ model in LPA to order $\rho^2$
    using separated cutoff scales. Due to the complete decoupling of
    both sectors, no dependence on the relative cutoff scale $c$ can
    be observed. Middle panel: BFP critical exponents of the same
    model in LPA to order $\rho^4$. Here, the two sectors are coupled
    showing a severe dependence on $c$. Right panel: Close-up of the
    largest BFP critical exponent $y_1$ of the
    $\text{O}(2)\oplus\text{O}(1)$-model in LPA to order $\rho^4$
    exhibiting the limiting cases $c \ll 1$ and $c \gg 1$ where one of
    the sectors is effectively suppressed and the critical behaviour
    is described by the Wilson-Fisher FP of the other sector.}
\label{relcDFP}
\end{figure*}
\noindent 
%


\subsection{Fixed Points and Relative Cutoff Scales}

\noindent In this section, we examine the effects that occur when
dealing with models whose sub-sectors are defined on separate cutoff
scales.  To that end, we investigate the $O(M)\oplus O(N)$ model using
flat regulators in the two sectors, however, with separated cutoff
scales
\begin{align}
  \nonumber r_\phi(y)&=\left(\frac{1}{y}-1\right)\theta(1-y)\,,\\[1ex] 
  r_\chi(y)&=r_\phi\left(\frac{y}{c}\right)=\left(\frac{c}{y}-1\right)
  \theta(1-\frac{y}{c})\label{eq:cutoff02}\,.
\end{align}
Consequently, for a generic $c\neq 1$ the fluctuations of one field
are integrated out earlier than the fluctuations of the other: The
second regulator has a built-in shift of all scales $k^2\rightarrow
c\,k^2, \Lambda^2 \rightarrow c\,\Lambda^2$.  For $c<1$ this leads to
a suppression of the $\chi$ sector and the RG flow does not experience
any $\chi$~fluctuations.  Inversely, $c>1$ suppresses the $\phi$
sector in a similar way. Only for $c=1$ the physical cutoff scales are
identical. Note that the statement about identical physical cutoff
scales $k_{\text{\tiny{phys}}}(\phi)=k_{\text{\tiny{phys}}}(\chi)$ is
only trivial in the present case where $\phi$ and $\chi$ are both
scalar fields with the same dispersion and interactions. In the
general case it is non-trivial to identify the relative cutoff scales
and where the different representations of the optimisation may
pay-off in particular.

The threshold functions for this choice of regulators with separate
cutoff scales can be found in App.~\ref{app:thresh}.  Now, we discuss
the dependence of critical exponents of different fixed points on
changes of the relative cutoff scale.


\paragraph*{DFP Critical Exponents} At the DFP the fields $\phi$ and
$\chi$ decouple.  Introducing a scale-shifted regulator in a single
$O(N)$-model, does not induce a difference in the critical exponents,
since every momentum can simply be rescaled.  The results of our
investigation for the DFP in the $O(2)\oplus O(1)$ model are displayed
in the left panel of Fig.~\ref{relcDFP}, confirming the previous
statement.  The critical exponents for all values of $c$ can be
identified with the Wilson-Fisher critical exponents
Tab.~\ref{relcCritExp} at the corresponding order of the truncation.

\paragraph*{IFP Critical Exponents} Introducing $c\neq 1$ revokes the
symmetry between $\phi$ and $\chi$, destroying the key property of
this fixed point.  Therefore, we will not further investigate this
fixed point in the present context. However, we note that this is an
illustrative example for how a unphysical regulator choice can lead to
artificial regulator dependencies of FRG predictions.

\paragraph*{BFP Critical Exponents} New insights can be gained by
looking at the $c$ dependence of the BFP where both sub-sectors are
explicitly coupled.  The critical exponents of the BFP clearly exhibit
a severe dependence on the relative scale factor $c$, see middle and
right panel of Fig.~\ref{relcDFP} for the example of the $O(2)\oplus
O(1)$ model.  In fact, the largest critical exponent, $y_1$, of the
BFP tends to the WF critical exponents of one of the single
sub-sectors as the other one is suppressed by a large relative cutoff
scale.  Explicitly, for $c \ll 1$, the critical exponent $y_1$ of the
$O(2)\oplus O(1)$ BFP approaches the value of the WF fixed point of
the $O(2)$ model.  Analogously, for $c\gg1$, $y_1$ approaches the WF
fixed point of $O(1)$ model.

We assert that in systems with various field degrees of freedom, the
choice of their relative cutoff scales has a severe impact on the
described physics in LPA due to the missing momentum dependences. The
change of the ratio $c$ of the cutoff scales induces a change of
universality classes. In the present simple $O(M)\oplus O(N)$ models
the generic choice is $c=1$ as the fields involved have identical
dispersions and interactions. We also emphasise again, that in more
complicated systems with different sectors, and in particular
fermion-boson systems, there is no clear \emph{a priori} criterion for
a suitable choice of regulators and their relative cutoff scales, see
also the following section.  In the inset of the right panel of
Fig.~\ref{relcDFP}, we further show that the value of the critical
exponent $y_1$ at $c=1$ is not singled out as a local extremum of the
critical exponent $y_1(c)$.  We suggest that a control of this issue
in multi-field models can be gained by the practical optimisation
procedure presented in Sec.~\ref{sec:opt} which, however, is beyond
the scope of the present work.


\subsection{Truncation Dependence}

\noindent The integrability condition \eq{eq:intcond} fails as we
truncate the effective action. In turn, this suggests that the
regulator dependence of the results becomes weaker when the level of
truncation is increased.  In this section we examine the dependence of
the BFP critical exponent $y_1$ on the relative cutoff scale factor
$c$ as a function of the level of truncation.  We focus on the
$O(2)\oplus O(2)$ model and compare different orders of the LPA, i.e.,
to the orders between $\rho^2$ up to $\rho^{6}$.

\begin{figure}[t!]
  \centering
  \includegraphics[width=0.85\columnwidth]{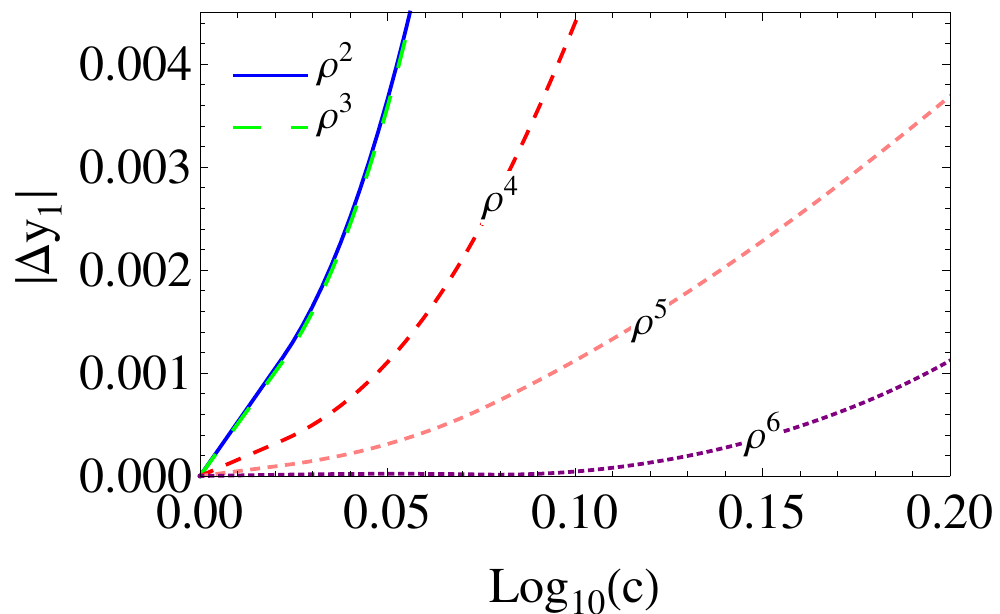}
  \caption{Deviation from the critical exponent $y_1$ in the
    $O(2)\oplus O(2)$ model in LPA from order $\rho^2$ to order
    $\rho^6$ as a function of the relative cutoff scale $c$. We
    clearly observe a flattening of the curves for higher orders of
    the truncation.}
\label{CritExpStababs}
\end{figure}

Fig.~\ref{CritExpStababs} shows the deviation of $y_1$ at a given
$c>1$ from the value at $c=1$ weighted by its difference to the
limiting case of the corresponding $O(2)$ critical exponent
\begin{align}\label{eq:absweight}
|\Delta y_1 |=\left| \frac{y_1(c)-y_1(1)}{y_1(1)-y_{1,\text{O(2)}}} \right|\,.
\end{align}
We see that for increasing order in the LPA from $\rho^2$ up to
$\rho^{6}$, the dependence of $y_1$ on the relative cutoff scale $c$
becomes weaker and weaker as suggested by the consecutive flattening
of the curves.  We conclude that a better truncation is more robust to
uneducated regulator choices or in other words a low-level truncation
requires a more sophisticated choice of the regulator scheme.

%


\section{Non-Relativistic Fermion-Boson Models}\label{sec:richard}

\noindent In the last Section \ref{sec:multifield}, we have discussed
the question of relative cutoff scales in a simple scalar model with
identical dispersions and interactions for the different fields. In
the present Section we discuss relative cutoff scales and the impact
of the shape dependence of regulators in a more complicated situation
of a non-relativistic Yukawa system describing fermionic atoms and
molecules. In contradistinction to the LPA approximation used in the
last Section we also take into account momentum and frequency
dependences of the propagators. Naturally, this does not fully cure
the lack of momentum dependences of the approximation and we expect a
modest regulator dependence of the corresponding results. Further
aspects in non-relativistic systems are $N$-body hierarchies which
lead to complete resummation schemes for three-, four- and $N$-body
systems, which has been worked out for fermionic three-, and four-body
cases for various systems, \cite{Diehl2008,Moroz2009,Schmidt2010,Birse2013,Tanizaki2013,Avila2015}. This gives us further access
for an assessment of the regulator dependence of our results.

The FRG has found multiple applications in the study of
non-relativistic systems, ranging from few-
\cite{Diehl2008,Moroz2009,Floerchinger2011} to many-body problems
\cite{Dupuis2007,Diehl2007,Floerchinger2008,Floerchinger2008b,schm2011,Rancon2011}.
In a prototypical scenario in condensed matter physics, fermionic and
bosonic degrees of freedom interact with each other.  Such situations
occur for instance in models which describe the formation of molecules
from atoms.  Other examples include the interaction of electrons with
collective excitations, such as phonons or magnons. In addition to the
question of equivalent cutoff scales of fermions and bosons, similarly
to the coupled $O(N)$ models discussed in Section
\ref{sec:multifield}, the exact fermionic $N$-body hierarchies
present an additional challenge for the mixed non-relativistic
bosonic-fermionic system within the evaluation of FRG flows. Optimal
approximations have to take into account these exact hierarchies in
addition to taking care of the momentum transfer. If both properties
cannot be rescued in a given approximation, it is a priori not clear,
in which order the various fields have to be integrated out for
optimal results. Hence, in this situation, the question of the optimal
ratio of cutoff scales is even more complicated as in the bosonic
example treated in Section \ref{sec:multifield}.

In the present Section we do not aim at a full resolution of this
intricate question, but rather highlight the ensuing difficulties. We
provide an analysis of the regulator dependences arising in a system
consisting of a single impurity immersed in a non-relativistic Fermi
sea of atoms at zero temperature. In this so-called Fermi polaron
problem \cite{prok2008,SchirotzekPRL09,schm2012,Kohs2012} the
interaction of the impurity $\psi_\downarrow$ with the fermions
$\psi_\uparrow$ in the Fermi sea is determined by the exchange of a
molecular field $\phi$ which represents a bound state of the
$\downarrow$-impurity with one of the medium $\uparrow$-atoms. The
system is described by the action
\begin{align}\label{pol3d.FBaction}
  S =& \int_{\bf{ x}, \tau} \Big\{\sum_{\sigma=\uparrow,\downarrow}
  \psi_\sigma^*[\partial_\tau - \Delta - \mu_\sigma]\psi_\sigma
  \nonumber\\
  &+\phi^*[\partial_\tau - \Delta/2 + \nu_\phi]\phi
  +h(\psi_\uparrow^*\psi_\downarrow^*\phi+h.c.)\Big\},
\end{align}
where $\int_{\bf{ x}, \tau}=\int d^3x d\tau$, $\Delta$ is the Laplace
operator and we suppressed the arguments $\bf{x}, \, \tau$ of the
fields. Furthermore the Grassmann-valued, fermionic fields
$\psi_\uparrow$ and $\psi_\downarrow$ represent $\uparrow$- and
$\downarrow$-spin fermions of equal mass $m$. Note that we work in
units $\hbar=2m=1$ and $\sigma=(\uparrow,\downarrow)$.  The associated
chemical potentials $\mu_\sigma$ are adjusted such that the
$\uparrow$-fermions have a finite density $n_\uparrow =
k_F^3/(6\pi^2)$, with $k_F$ the Fermi momentum, while there is only a
single $\downarrow$-atom. In this limit the action
Eq.~\eqref{pol3d.FBaction} describes the problem of a single impurity
immersed in a Fermi sea.  The detuning $\nu_\phi$, together with the
coupling $h$ determines the interaction strength between the
$\downarrow$- and $\uparrow$-atoms which is mediated by the exchange
of the field $\phi$.

The impurity is dressed by fluctuations in the fermionic
background. It becomes a quasi-particle, the Fermi polaron, which is
characterised by particle-like properties such as an energy $E_p$, and
a quasi-particle weight $Z_p$.  The quasiparticle properties depend on
the interaction between the impurity and the Fermi gas.  Due to the
presence of bound states, this interaction cannot be described within
perturbation theory and requires non-perturbative
approximations. Hence, it presents an ideal testing ground for methods
such as the FRG.

In the following the prediction of $E_p$ will serve as our observable
to study the regulator dependences occurring in the RG evaluation of
the model Eq.~\eqref{pol3d.FBaction}. The Fermi polaron problem is
particularly interesting for our study since accurate numerical
predictions for various quantities exist based on a bold diagrammatic
Monte Carlo scheme~\cite{prok2008}. For instance at unitarity, where
the infrared scattering amplitude at zero scattering momentum -- given
by the scattering length $a_s$ -- diverges, $k_Fa_s\to\infty$, the
ground state energy is predicted to approach the value of
$E_p=-0.615\, \epsilon_F$ \cite{prok2008}, with $\epsilon_F$ the Fermi
energy.

We note that also other non-relativistic systems of coupled bosons and
fermions are described by Eq.~\eqref{pol3d.FBaction}. For instance for
a chemical potential $\mu_\downarrow>-E_p$ the system exhibits the
BEC-BCS crossover at low temperature as interactions are varied
\cite{haus2007,bloc2008,zwerger2011}. This crossover has been studied
extensively by FRG methods
\cite{Diehl:2005ae,Diehl:2005an,Diehl2007,Floerchinger2008,Bartosch2009}.


\subsection{Truncation and Flow Equations}

\noindent In the following we will solve the FRG flow equation~\eqref{eq:flow}
for the truncation of the effective action
\begin{eqnarray}
  \label{pol.fulltrunc}
  \Gamma_k&=&\int_{{\bf p},\omega}\Big\{ \psi^*_\uparrow[-i\omega
  +{\bf p}^2-\mu_\uparrow]\psi_\uparrow+\psi^*_\downarrow
  G_{\downarrow,k}^{-1}(\omega,{\bf p})\psi_\downarrow \nonumber \\&+&\phi^* 
  G_{\phi,k}^{-1}(\omega,{\bf p})\phi\Big\}
  +\int_{\vec x,\tau}
  h (\psi_\uparrow^*\psi_\downarrow^*\phi+h.c.), 
\end{eqnarray}
where $\int_{\bf p, \omega}=\int \frac{d^3p}{(2\pi)^3} \int
\frac{d\omega}{2\pi}$. In this truncation the only RG scale $k$
dependent quantities are $G_{\downarrow,k}$ and $G_{\phi,k}$. While in
previous work the flow of fully momentum dependent propagators
$G_{\downarrow,k}$ and $G_{\phi,k}$ has been considered
\cite{schm2011}, we study here the regulator dependences arising in a
derivative expansion where
\begin{eqnarray}\label{pol.derivprops}
  G_{\downarrow,k}^{-1}(\omega,\bf p)&=&S_\downarrow[-i \omega+{\bf p}^2]+
  m_\downarrow^2,\nonumber\\[2ex]
  G_{\phi,k}^{-1}(\omega,\bf p)&=&S_\phi[-i \omega+{\bf p}^2/2]+m_\phi^2\,,
\end{eqnarray}
with scale-dependent wave function renormalisations
$S_\downarrow,S_\phi$. It has been shown in \cite{Helmboldt:2014iya}
for relativisitic Yukawa systems that this approximation of the full
frequency- and momentum-dependence already captures well the full
dependence. We expect this to be also the case in the present
non-relativistic case. The RG scale dependent coupling constants
$m^2_\downarrow$ and $m^2_\phi$ are related to the flowing static
self-energies $\Sigma_{\downarrow,\phi}(0,\mathbf{0})$,
e.g.~$m_\downarrow^2=-\mu_\downarrow-\Sigma_\downarrow(0,\bf{ 0})$.
In the impurity problem the majority fermions are not renormalized,
$S_\uparrow=1$, and the density of the Fermi sea is determined by the
chemical potential $\mu_\uparrow=\varepsilon_F=k_F^2$. In summary,
from this truncation, we obtain the four flow equations,
cf.~App.~\ref{app:polaron},
\begin{eqnarray}
  \partial_t m_\phi^2&=&\frac{h^2}{2\pi^2}\int_{k_F}^\infty dp\frac{p^2(
\partial_t R_\downarrow+ S_\downarrow \partial_t R_\uparrow)}{[m_\downarrow^2
+R_\downarrow+S_\downarrow(2p^2-\mu_\uparrow+R_\uparrow)]^2}\nonumber\\[2ex]
\partial_t S_\phi&=&-\frac{ h^2}{\pi^2}\int_{k_F}^\infty dp\frac{S_\downarrow 
p^2( \partial_t R_\downarrow+ S_\downarrow \partial_t R_\uparrow)}{[
m_\downarrow^2+R_\downarrow+S_\downarrow(2p^2-\mu_\uparrow+R_\uparrow)]^3}\nonumber\\[2ex]
\partial_t m_\downarrow^2&=&\frac{h^2}{2\pi^2}\int^{k_F}_0 dp
\frac{p^2(\partial_t R_\phi- S_\phi \partial_t R_\uparrow)}{[m_\phi^2
  +R_\phi-S_\phi(p^2/2-\mu_\uparrow+R_\uparrow)]^2}\nonumber\\[2ex]
\partial_t S_\downarrow^2&=&-\frac{h^2}{\pi^2}\int^{k_F}_0 dp
\frac{S_\phi p^2(\partial_t R_\phi- S_\phi \partial_t R_\uparrow)}{[m_\phi^2
+R_\phi-S_\phi(p^2/2-\mu_\uparrow+R_\uparrow)]^3}.\nonumber\\
\end{eqnarray}
We study the dependence of the predictions from the FRG using a
continuous set of regulators $R_{\downarrow,\uparrow}$ and $R_\phi$
which are dependent on various parameters. We choose
\begin{eqnarray}
  R_\phi^k(p)&=&c_\phi \frac{S_\phi k^2}{2}(a_\phi-b_\phi y)
  \frac{y^{n_\phi}}{e^{y^{n_\phi}} -1}\nonumber\\
  R_\downarrow^k(p)&=&c_\downarrow S_\downarrow k^2 (a_\downarrow-b_\downarrow y)
  \frac{y^{n_\downarrow}}{e^{y^{n_\downarrow}} -1}\nonumber\\
  R_\uparrow^k(p)&=&c_\uparrow S_\uparrow k^2 (a_\uparrow-b_\uparrow y)
  \sigma(p^2-\mu_\uparrow)\frac{y^{n_\uparrow}}{e^{y^{n_\uparrow}} -1}\label{polregulators}
\end{eqnarray}
where $y\equiv p^2/(c_i k^2)$ and $\sigma(x)=1,\,(-1)$ for $x>0$
$(x\leq 1)$ for the impurity $\psi_\downarrow$ and boson field $\phi$.
These regulators are similar to the regulators studied in the
relativistic models in the previous sections,
cf. Tab.~\ref{tregs}. Note however that for the bath fermions the pole
structure due to the Fermi surface has to be accounted for so that
here $y\equiv (|{\bf p}^2-\mu_\uparrow|)/(c_\downarrow k^2)$. Similar
to the definitions used in Section \ref{sec:multifield}, the
parameters $c_{i}$ ($i=\phi,\uparrow,\downarrow$) allow for the study
of changing the relative scales at which the various field are
integrated out, while the other parameters allow for deformations of
the regulator shape, cf.~Fig.~\ref{fig:propnorm}.


\subsection{Initial Conditions}

\noindent As discussed in Section \ref{sec.uvreg}, first the initial
values at the UV scale $k=\Lambda$ have to be set. The initial value
of $m_\phi^2$ is determined by the interaction strength between the
impurity and fermions in the Fermi sea. This interaction strength is
given by the low-energy scattering length $a_s$. The latter is
determined by the evaluation of the tree-level exchange of the
molecule field $\phi$ in the two-body problem where
$\mu_{\downarrow,\uparrow}=0$. This results in the initial value
\begin{align}
  m_{\phi\Lambda}^2=-\frac{h^2}{8\pi
    a_s}-\frac{h^2}{2\pi^2}\int_0^\infty dp\,
  p^2\left[\frac{1}{2p^2+R^\Lambda_\uparrow+
R_\downarrow^\Lambda}-\frac{1}{2p^2}\right]\,.\nonumber
\end{align}
For large cutoff scales $\Lambda$ this implies the scaling
$m_{\phi\Lambda}^2\sim\mu(r)h^2\Lambda$ with $\mu(r)$ being a
regulator dependent number. It is the non-relativistic equivalent to
the UV scaling discussed earlier, cf. Eq.~\eqref{eq:mUV}.
Furthermore, while the initial value of $S_\downarrow$ is determined
by its classical value $S_{\downarrow\Lambda}=1$, we choose $S_\phi=0$
so that the bosonic field becomes a pure auxiliary field.

The Fermi momentum allows to define the dimensionless interaction
parameter $1/(k_Fa_s)$. In the following we work in units where
$k_F=1$. Finally the initial value of $m_{\downarrow\Lambda}^2$ has to
be chosen such that the self-energy acquired by the impurity leads to
the fulfillment of the infrared condition $m^2_{\downarrow
  k=0}=0$. This condition ensures that the system is just on the verge
of occupying a finite number of impurity atoms, which is the defining
property of the impurity problem. This condition implies
$E_p=-m_{\downarrow\Lambda}^2=\mu_\downarrow$ \cite{schm2011}.


\subsection{Regulator Dependencies}

\begin{figure}[t]
  \centering
  \includegraphics[width=0.92\columnwidth]{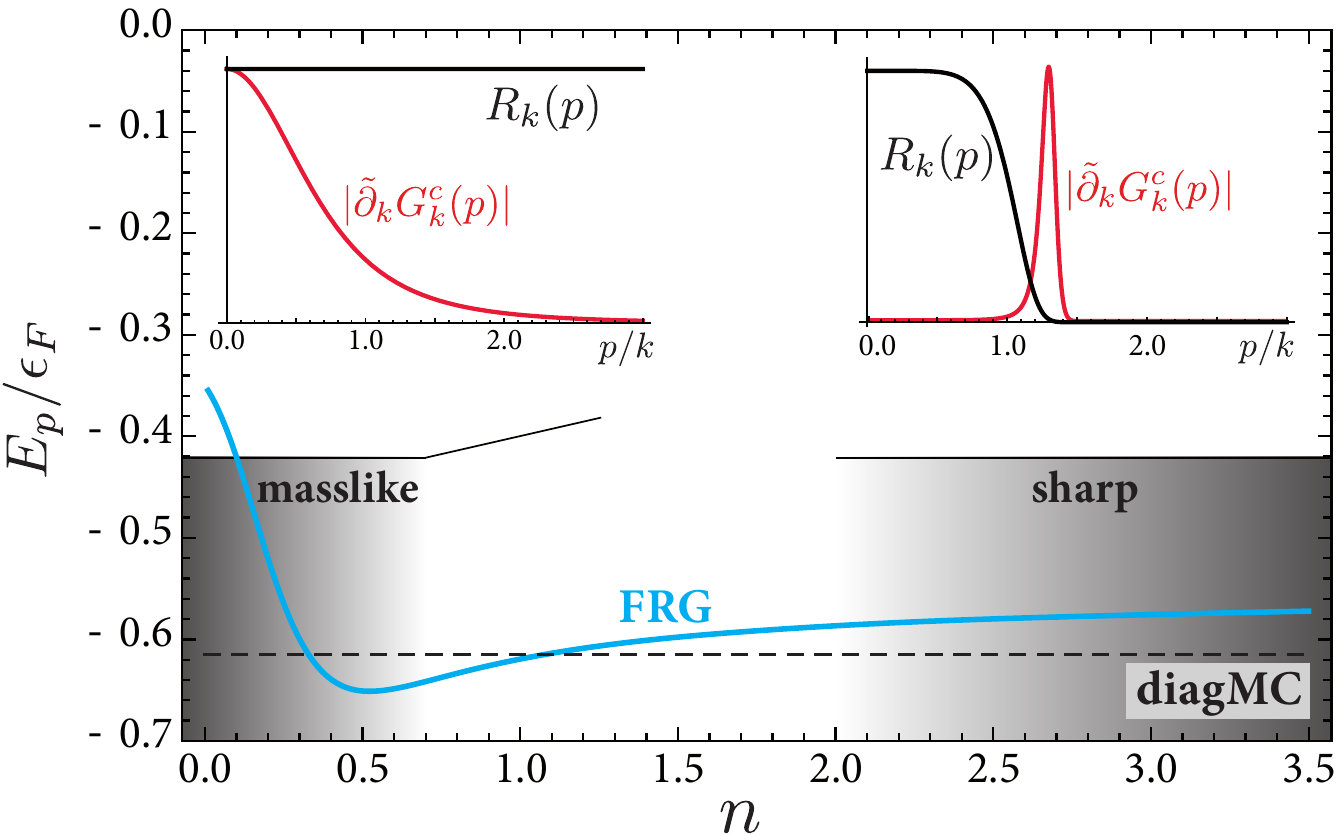}
  \caption{Dependence of the polaron energy $E_p/\epsilon_F$ on the
    regulator shape. The FRG results (blue) are shown for a crossover
    from a $k^2$ to a sharp regulator by changing the exponent
    $n\equiv n_i$ in all regulators at constant prefactors $a_i=100$
    and $b=0$. The exact result from diagramatic Monte Carlo
    \cite{prok2008} is shown as dashed black line. The insets
    illustrate the structure of the regulator $R_\downarrow(p)$ and
    single scale propagator $\tilde\partial_k G_{\downarrow,k}^c = - 
    (G_{\downarrow,k}^c)^2 \partial_kR_k$ in dependence of momentum
    $p$.}\label{Fig.Pol.Dep1}
\end{figure}

\paragraph*{Shape dependence.}
First we study the dependence of the results on the shape of the
regulators when integrating out the bosonic and fermionic fields
synchronously for the
choice $c_\phi=c_\downarrow=c_\uparrow=1$. Specifically, we monitor the
energy of the polaron, $E_p$. In Fig.~\ref{Fig.Pol.Dep1} we show the
result for the polaron energy at unitary interactions, $k_F a=\infty$,
as a function of the shape parameters $n\equiv
n_\uparrow=n_\downarrow=n_\phi$ (blue line). We have studied such a
variation previously in Section \ref{sec:pracopt} in the context of a
relativistic $\varphi^4$ theory.  Here, we choose $b_i=0$ and
$a_i=100$ so that the regulator interpolates between a masslike $k^2$
and a sharp regulator.  The black dashed line corresponds to the
result obtained from diagramatic Monte Carlo, $E_p/\epsilon_F\approx
-0.615$ \cite{prok2008}. We note that a non-selfconsistent T-matrix
approximation yields the result $E_p/\epsilon_F\approx -0.607$
\cite{comb2007}. This approximation (leading order $1/N$ expansion)
corresponds to the sequence where first the dimer selfenergy is
evaluated and then inserted into the self-energy of the impurity. 

The result from the FRG calculation is shown as blue line. We 
observe a strong regulator shape dependence for small values of $n$
(masslike regulator) as, here, the regulator leads to a non-local
integration of field modes in momentum space. 

In contrast, for $n\to\infty$ (sharp regulator), the regulator becomes
very local in momentum, and the results show only a small shape
dependence. In the flows, the single scale propagator
$\tilde\partial_k G_k^c=- (G_{k}^c)^2 \partial_kR_k$ determines the
locality of the regulator in momentum space which is illustrated in
the insets in Fig.~\ref{Fig.Pol.Dep1}. Here, we show the form of the
single scale propagator as well as $R_\downarrow(p=|{\bf p}|)$
evaluated at zero frequency $\omega=0$. We emphasise that momentum
locality of the loop integration is but one of the important
conditions for the optimisation. Regularity of the flow is a further
important one, and the sharp cutoff fails in this respect. Indeed, it
is the latter property which is crucial for critical exponents. 

For $n\to 0 $ the non-local structure of field integration leads to a
great sensitivity of the RG flow of $\Gamma_k$ in theory space and
hence a large sensitivity to the truncation chosen.  This also finds
an interpretation in terms of physics: due to the non-local structure
of $R_k(p)$ the RG flow does not separate between the few-body
(vacuum) physics at large momenta on the one hand and on the other
hand the emergence of correction to the vacuum flow due to finite
density at small momenta. Such an unphysical mixture of physically
vastly separated energy scales leads to an artificially strong
dependence of the FRG results. We indicate the regime of artificial
non-local, non-physical regulator choices by the grey shaded area in
Fig.~\ref{Fig.Pol.Dep1}. The results illustrate the significance of
the statement that for local truncations non-local cutoffs are a
particularly bad choice of regularisation of RG flows. Instead general
RG flows should be kept sufficiently local. However, also an extremely
local regulator such as the sharp regulator is not desired as it
prevents an interference of closely related momentum/length scales; it
lacks regularity. We have indicated this regime as shaded area for $n>2$
where the single scale propagator becomes strictly peaked and
interference of close-by momentum scales is heavily suppressed. The
non-shaded regime corresponds to regulator choices which satisfy the
criterion of sufficient interference of momentum scales while still
avoiding an unphysical, non-local flow.

In summary, both extrem choices lack crucial
properties of optimised FRG flows. This is also reflected in the fact
that both limits do not do well in the optimisation criterium in its
representations \eq{eq:minbarG} and \eq{eq:Optlength}. Indeed, the
combination of a sharp cutoff and a mass cutoff gives the worst result
within the optimisation as it combines both failures, momentum
nonlocality and lack of regularity.\\[-1ex]

\begin{figure}[t!]
  \centering
  \includegraphics[width=0.92\columnwidth]{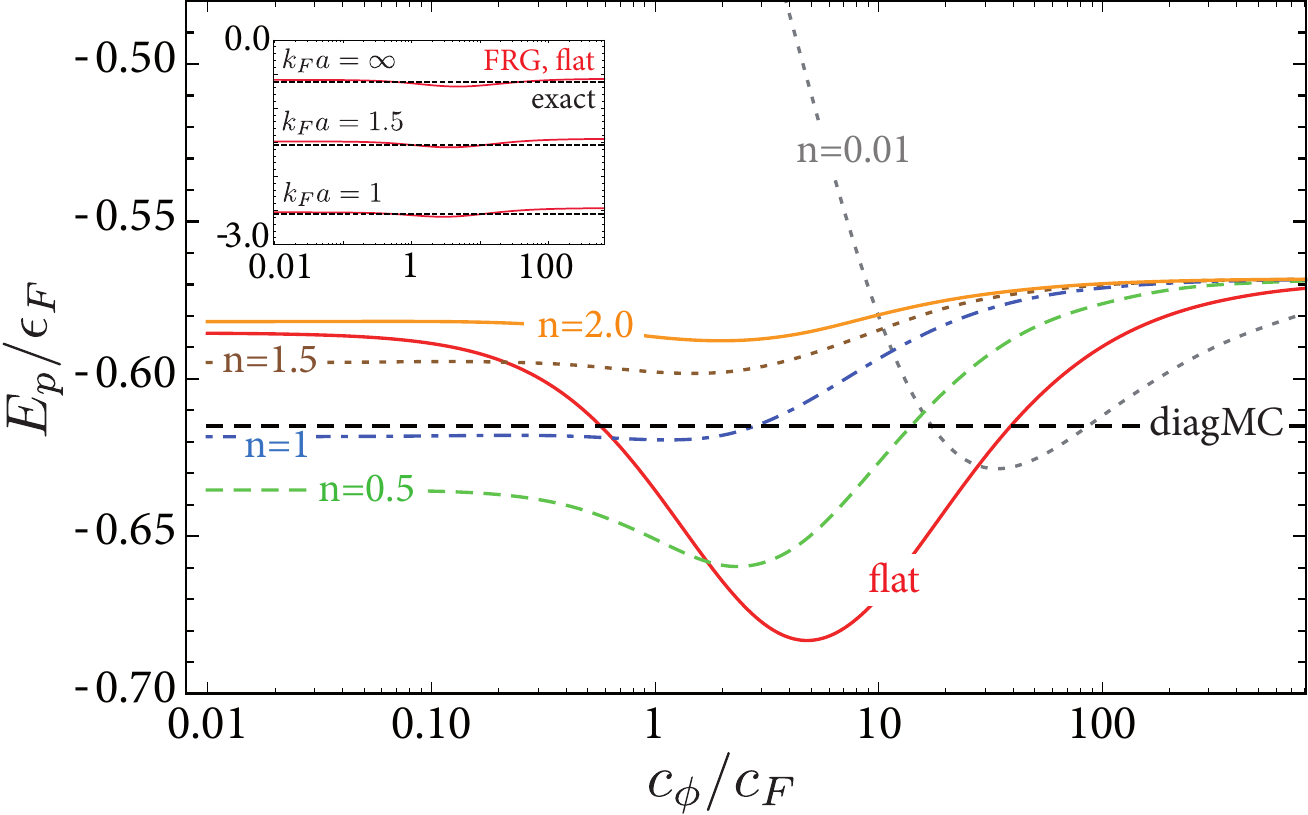}
  \caption{Dependence of the polaron energy $E_p/\epsilon_F$ on the
    relative bosonic-fermionic cutoff scale for various choices of the
    regulator shape. Results are shown for parameters as employed in
    Fig.~\ref{Fig.Pol.Dep1} and also for the flat (Litim) regulator,
    cf.~Table~\ref{tregs}.  The inset shows the relative cutoff scale
    dependence for the flat regulator for various interaction
    strengths $1/k_Fa\neq 0$.}\label{Fig.Pol.Dep2}
\end{figure}


\paragraph*{Dependence on relative cutoff scale.} Next, we investigate
the dependence of the results on the relative scale at which the fermionic
and bosonic degrees of freedom are integrated out. As in Section
\ref{sec:multifield}, this is achieved by changing the parameter
$c_\phi$ relative to the choice of $c_F\equiv c_\downarrow =
c_\uparrow$ in Eq.~\eqref{polregulators}. The result is shown in
Fig.~\ref{Fig.Pol.Dep2} where we choose the same regulator shape
parameters as in Fig.~\ref{Fig.Pol.Dep1}. Also we show the result for
a flat regulator choice (red curves).

For $c_\phi/c_F\to\infty$ the flow is equivalent to a purely fermionic
flow since here the auxiliary bosons are integrated out only in the
last step of the RG flow. In this last step the fermions are not
subject to an RG gap $R_{\sigma}$ anymore. In consequence, since the
flow of the boson propagator is solely dependent on fermions, its
self-energy $\Sigma_\phi$ reached its final RG value already before
this last RG step is taken. This leads to results which are
independent of the regulator shape, $E_p/\epsilon_F\to -0.57$, and
which is the result obtained from an leading order $1/N$
expansion~\cite{Enss2012} within our truncation for the momentum
dependence of the bosonic propagator $P_\phi$.

Contrary, for $c_\phi/c_F\to0$ the bosonic field is integrated out
first. The flow of the boson propagator, being only a functional of
the fermionic Green's functions, is then completely suppressed in the
first stage of the RG flow. This stage correspondingly
amounts to a mere reversion of the introduction of the bosonic field
$\phi$ as an auxiliary degree of freedom mediating the atom-atom
interaction. Since $P_\phi$ cannot acquire any momentum
dependence in this step, the resulting -- now purely fermionic --
theory has a truncation with a completely momentum independent
coupling constant $\lambda\sim -h^2/m_{\phi,\Lambda}^2$.

Such a truncation of the effective flowing action $\Gamma_k$ is of
course a very poor one so that strong regulator dependencies are
expected as also observed in Fig.~\ref{Fig.Pol.Dep2}. This result also
represents an example supporting the discussion given in Section
\ref{sec:init}, cf.~also Fig.~\ref{fig:sflow}: by choosing a poor
truncation the flow is particularly sensitive to its path in theory
space and hence can lead to strong regulator dependences of infrared
quantities. Furthermore this result reflects the observation that the
integrability condition \eqref{eq:intcond} is more severely violated
when the effective action is truncated to a larger degree (here by
loosing the momentum dependence of interactions altogether).

Having shown that the regulator dependences in the two extreme limits
for $c_\phi/c_F$ can be understood in simple terms we now turn to
the intermediate regime where the bosonic and fermionic fields are
integrated out synchronously. In this regime we observe a variation of
the result for $E_p/\epsilon_F$ on the order of $\pm 10\,\%$, with the
exact result $E_p/\epsilon_F=-0.615$ being in the vicinity of the
predicted result by the FRG.

We also show the result when applying the flat regulator (red line)
which shows similar variations with the relative cutoff scale. Our
results indicate that within the regime of an `informed regulator
choice', indicated by the non-shaded region in
Fig.~\ref{Fig.Pol.Dep1}, regulator dependences in FRG flows might
allow for determining an error estimate on its own predictions.


\section{Conclusion}

\noindent In this work, we have presented a systematic investigation
of the impact of different regulator choices on
renormalisation group flows in given approximation schemes.  To this
end, we studied the functional RG which is based on the scale-dependent effective
action.  As an important aspect, this exact flow equation clearly
exhibits the role played by the regulator within the RG, see
\eq{eq:flow}, as it is directly proportional to its scale
derivative. This already indicates the need for a thorough
understanding of regulator dependencies.  Such an understanding is not
only important to the functional RG, in particular, but, more
comprehensively, extends to the analysis of approximations schemes in
the renormalisation group framework in general. 

Here, we focused on three key aspects of how the regulator choice
affects RG results: First, we discussed how the choice of a specific
regulator influences FRG flows by integrating over flow trajectories
along closed loops in the space of action functionals varying both,
the regulator scale and its shape function.  For these flows we have
discussed an integrability condition, \cite{Pawlowski:2005xe}, which
is violated in the presence of truncations.  Consequently, an educated
regulator choice is mandatory to extract the best possible results
from the RG in a given truncation.  To this end we have extended the
work on functional renormalisation in \cite{Pawlowski:2005xe}. For the
construction of such an optimised regulator, we have introduced the
definition of the length of a RG trajectory which is minimal for an
optimised regulator. This provides a pragmatic optimisation procedure
which at the example of a single scalar field yields the flat 
regulator as a unique and analytical solution. A comparison of the
lengths of these trajectories can also be set up straightforwardly in
more complex models in order to identify optimised regulators. We
leave explicit applications of this procedure for future work.

As a second aspect, we have investigated systems with two field
degrees of freedom which both have to be regularised.  Here a choice
of relative cutoff scales is required. In given momentum-independent
approximations this choice has a severe impact on the RG results and,
hence, for the described physics.  At the example of relativistic
bosonic two-field models, we have discussed the consequences of a
variation of the relative cutoff scales as well as its truncation
dependence.  We have shown that a crossover between different
universality classes can be induced, triggered by the
regulator-dependence of physical parameters in truncated flows.  This
entails that the relative cutoff scale has to be chosen carefully for
a reliable description of a physical system in a given approximation.
A controlled approach toward devising an optimised choice of relative
cutoff scales can be provided by our optimisation procedure.

Third, we also have exhibited corresponding dependencies on relative
cutoff scales and regulator shapes in non-relativistic continuum
models of coupled fermionic and bosonic fields.  At the example of the
Fermi polaron problem in three spatial dimensions, we have illustrated
such dependences and showed how to interpret them in physical terms.
We suggested that, in the regime of an informed regulator choice,
regulator dependences in FRG flows can provide error estimates. This
has been discussed here at the example of a coupled non-relativistic
many-body model. It will be interesting to investigate these
capabilities further in more elaborate many-body models. Finally, it
is of great interest to extend the functional optimisation framework
layed out here and in \cite{Pawlowski:2005xe} to an approach for
general systematic error estimates in the functional RG.

\paragraph*{Acknowledgements}
We thank N.~Prokofiev and B.~Svistunov for providing
their diagMC data. Further, we thank A. Rodigast and
I. Boettcher for discussions. R.S. was supported by the NSF through a
grant for the Institute for Theoretical Atomic, Molecular, and Optical
Physics at Harvard University and the Smithsonian Astrophysical
Observatory.  S.W. acknowledges support by the Heidelberg Graduate
School of Fundamental Physics. This work is supported by the Helmholtz
Alliance HA216/EMMI, and the grant ERC-AdG-290623.


\appendix


\section{Effective Action and RG Transformations}
\label{app:rgtrafo}

\noindent Note also, that the above renormalisation scheme dependence carries
over to the full quantum effective action
$\Gamma[\phi]=\Gamma_{k=0}[\phi]$. It satisfies the standard
homogenous RG equation
\begin{align}\label{eq:RGmu} 
  s \0{d\Gamma_{k=0}[\phi]}{ds} =0\,,
\end{align} 
\eq{eq:RGmu} is non-trivially achieved as all correlation functions
$\Gamma^{(n)}$ transform according to the anomalous dimension of the
fields,
\begin{align}\label{eq:RGGn} 
  \left( \partial_s +n\gamma_\phi\right)\Gamma^{(n)}=0\,,\quad {\rm
    with}\quad \gamma_\phi \phi = s \0{d \phi}{ds}\,.
\end{align} 
For the purpose of the present work the RG-transformations of the full
effective action are not relevant. Hence, from now we shall identify
observables that are identical up to RG transformation of the
underlying theory.  Note however, that this identification does not
remove the relevant UV scaling carried by \eq{eq:mUV}. 


\section{General One-Parameter Flows}
\label{app:onparam}

\noindent We have introduced one-parameter flows, referring to general
changes of the cutoff scale $k$ with $k(s)$, changes of the shape of
the regulator, $r_s$ as well as reparametrisations of the theory.
The corresponding flow equation has the same
form as that for the $k$-flow in \eq{eq:flow}.
It reads
\begin{align}\label{eq:flows} 
  s\0{d\Gamma[\phi,R] }{ds} = \012 \Tr\,G[\phi,R] \left(\partial_s +2
    \gamma_\phi\right) R \,,
\end{align} 
where the total derivative w.r.t.\ $s$ also includes
reparamterisations of the fields with $d\phi/ds = \gamma_\phi \phi$
reflected in the term proportional to the anomalous dimension
$\gamma_\phi$ on the right hand side of \eq{eq:flows}.
The
representation of the total $s$ derivative similar to \eq{eq:repdt} is
simply given by
\begin{align}\label{eq:repds}
  \0{d}{ds} = \left( -\012 \Tr\, G[\phi,R]\,\left (\partial_s +2
      \gamma_\phi\right) R\, G[\phi,R]\0{\delta^2}{\delta\phi^2}
  \right) \,.
\end{align}
Note that \eq{eq:repds} has to vanish as an operator if it represents
a reparameterisation of the theory at hand, that is a standard
renormalisation group transformation in the presence of a regulator.
We infer, \cite{Pawlowski:2005xe},
\begin{align}\label{eq:sRGtrafo} 
\left( \partial_s +2      \gamma_\phi\right) R \stackrel{!}{=}0\,. 
\end{align}
\Eq{eq:sRGtrafo} entails that the regulator has to be transformed as a
two-point correlation function under RG-transformations in order to
fully reparameterise the theory. 


\section{Integrability Condition and Self-Consistency of Approximations}\label{app:intcond}


\noindent In case the integrability condition \eq{eq:closedloop} holds,
the flow necessarily has a (local) representation as a total
derivative w.r.t. $s$, and hence it can be written as a total
derivative of a diagrammatic representation. This entails that the
integrated flow has a diagrammatic representation in terms of full
vertices and propagators in the given approximation to the effective
action $\Gamma_k$. 

A simple example for such an approximation is perturbation theory: at
perturbative $n$-loop order the integrated flow simply reproduces
renormalised perturbation theory within a generalised
BPHZ-scheme. Note however that the ordering scheme is an expansion in
the fundamental coupling of the theory for both, the effective action
and the flow equation, rather than one in expansion coefficients of
the effective action such as the vertex expansion in terms of
$\Gamma_k^{(n)}$.  A more interesting example are 2PI-resummation
schemes such as 2PI perturbation theory or $1/N$-expansions: it has
been worked out how to implement renormalised versions of these
schemes in the FRG, see
\cite{Blaizot:2010zx,Carrington:2014lba}. Hence in this case
integrated flows provide renormalised perturbative or $1/N$
2PI-resummations, and the integrability condition \eq{eq:closedloop}
is satisfied to any order of such an expansion. Again we note that the
ordering scheme is an expansion in the fundamental coupling of the
theory or the number of fields for both, the effective action and the
flow equation. Similarly it is possible to find approximation schemes
that lead to renormalised solutions of Dyson-Schwinger equations. 

We emphasise that in both cases discussed above the flow operators
\eq{eq:repdt}, \eq{eq:repds} evaluated on the solution of the
effective action in the given approximation, does not satisfy the
integrability condition. In the case of $n$-loop perturbation theory
the flow operators \eq{eq:repdt}, \eq{eq:repds} then generates
$n$-loop FRG-resummed perturbation theory which fails to satisfy
\eq{eq:closedloop}. In the case of the $n$-loop 2PI approximation, the
flow operators then generate $n$-loop FRG-resummed 2PI perturbation
theory. To summarise, the violation of the integrability condition is a measure for the incompleteness, in terms of the full quantum theory, of fully
non-perturbative resummation schemes.


\section{Threshold Functions}\label{app:thresh}


\subsubsection*{Scalar Model}

\noindent The scalar model from Sec.~\ref{sec:scalarloops} requires
the threshold function $I(\omega)$ defined in \eq{eq:uthresh}. Here,
we explicitly give the analytical expressions for this integral for
the cases of the flat regulator $r_{\text{L}}$ and the sharp regulator
$r_{\text{sharp}}$.  For the flat regulator, we obtain $I(\omega)
\rightarrow I^{\text{(L)}}(\omega)$
\begin{align}
I^{\text{(L)}}(\omega)&=v_d\frac{4}{d}\frac{1}{1+\omega}\,.
\label{thflat}
\end{align}
\noindent Choosing the sharp regulator yields $I(\omega) \rightarrow
I^{\text{(sharp)}}(\omega)$
\begin{align}
I^{\text{(sharp)}}(\omega)&= -2v_d\log(1+\omega)\,.
\label{thsharp}
\end{align}


\subsubsection*{Two-Field-Model}

For the two-field models from Sec.~\ref{sec:multifield}, we have
introduced similar threshold functions reading
\begin{align}\label{eq46a}
  &I_{R,\phi}(\omega_\phi,\omega_\chi,\omega_{\phi \chi})=v_d\int_0^\infty y^{\frac{d}{2}+1}dy \ \big( -2 r^\prime_\phi(y)\big)\nonumber \\
  &\ \times\frac{y(1+
    r_\chi(y))+\omega_\chi}{(y(1+r_\phi(y))+\omega_\phi)(y(1+r_\chi(y))+\omega_\chi)-\omega_{\phi
      \chi}} \ ,
\end{align}
\begin{align}\label{eq46b}
  &I_{R,\chi}(\omega_\phi,\omega_\chi,\omega_{\phi \chi})=v_d\int_0^\infty y^{\frac{d}{2}+1}dy \ \big( -2 r^\prime_\chi(y)\big) \nonumber \\
  &\ \times\frac{y(1+
    r_\phi(y))+\omega_\phi}{(y(1+r_\phi(y))+\omega_\phi)(y(1+r_\chi(y))+\omega_\chi)-\omega_{\phi
      \chi}} \ ,
\end{align}
and
\begin{align}\label{eq46c}
&I_{G,i}(x)=v_d\int_0^\infty y^{\frac{d}{2}+1}dy \ \frac{-2 r^\prime_i(y)}{y(1+r_i(y))+x}\,,
\end{align}
with $i\in \{\phi,\chi\}$.


\subsubsection*{Two-Field Model \& Separate Cutoff Scales}
\label{arelcutoff}

\noindent For the discussion of the two-field model in
Sec.~\ref{sec:multifield}, we use the flat regulator functions with a
relative cutoff scale, as given in Eq.~\eq{eq:cutoff02}. With these shape functions, we obtain the threshold
functions for the Goldstone modes
\begin{align}
I_{G,\phi}(x)&=\frac{4 v_d }{d\, (x+1)}\ ,\quad
I_{G,\chi}(x)=\frac{4 v_d\, c^{d/2+1} }{d\, (x+c)}\,.
\end{align}
The threshold function including radial modes are given by the expressions
\begin{align}
  &I_{R,\phi}(\omega_\phi,\omega_\chi,\omega_{\phi\chi})=2v_d
  \theta(1-c)
  F_1(\omega_{\phi},\omega_{\chi},\omega_{\phi\chi}) \nonumber \\
  &\quad\quad\quad\ +\frac{4 v_d (c+\omega_{\chi}) \left(c^{d/2}
      \theta(1-c)+\theta(c-1) \right)}{d \left((\omega_{\phi}+1)
      (c+\omega_{\chi})-\omega_{\phi\chi}\right)}\,,\\
  &I_{R,\chi}(\omega_\phi,\omega_\chi,\omega_{\phi\chi})=-2c\,v_d
  \theta(c-1) F_2 (\omega_{\phi},\omega_{\chi},\omega_{\phi\chi})\nonumber \\
  &\quad\quad\quad\ +\frac{4 v_d (\omega_{\phi}+1)
    \left(c^{\frac{d}{2}+1} \theta(1-c) +c\,\theta(c-1)\right)}{d
    \left((\omega_{\phi}+1)
      (c+\omega_{\chi})-\omega_{\phi\chi}\right)} \label{eq51}\,,
\end{align}
where we have introduced the two integral functions
\begin{align}
  F_1(\omega_\phi,\omega_\chi,\omega_{\phi\chi})&=\int_c^1
  \textrm{d}y\frac{y^{\frac{d}{2}-1} (\omega_{\chi}+y)}{(\omega_{\phi}
    +1) (\omega_{\chi}+y)-\omega_{\phi\chi}}\,,\\
  F_2(\omega_\phi,\omega_\chi,\omega_{\phi\chi})&=\int_c^1
  \textrm{d}y\frac{y^{\frac{d}{2}-1}
    (\omega_{\phi}+y)}{(c+\omega_{\chi})
    (\omega_{\phi}+y)-\omega_{\phi\chi}}\,.
\end{align}
This completes our list of required threshold functions for the
two-field model with shape functions defined on separate cutoff
scales.


\section{Flow equations for the couplings}
\label{app:twofieldcouplings}

\subsubsection*{Scalar Model}

\noindent In the symmetric regime, where $\kappa=0$, we obtain the
flow of the coupling constants $\partial_t{\lambda}_{i}=
(\partial_t{u})^{(i)}|_{\rho=0}$ from the $i$th derivative of
$\partial_tu(\rho)$ with respect to $\rho$.

Analogously, in the symmetry broken regime, where $\lambda_1=0$ and
$\kappa >0$, we get
\begin{align}
  \partial_t{\kappa}=-\frac{(\partial_t
    u)'}{\lambda_{2}}\Big\rvert_{\rho=\kappa}\,,\quad \partial_t{\lambda}_{i\geq
    2}=(\partial_t {u})^{(i)}
  +u^{(i+1)}\partial_t{\kappa}\,\Big|_{\rho=\kappa}\,.\nonumber
\end{align}

\subsubsection*{Two-Field Model}

\noindent Projecting the flow equation on the definition of $u$ gives us the system of beta functions for the couplings
\begin{align}
\partial_t{\kappa}_{\phi}&=-\frac{{\lambda}_{02}(\partial_t{u})^{(1,0)}-{\lambda}_{11}(\partial_t{u})^{(0,1)}}{{\lambda}_{20}{\lambda}_{02}-{\lambda}_{11}^2}\Big\rvert_{\kappa_\phi,\kappa_\chi} \ ,\nonumber \\
\partial_t{\kappa}_{\chi}&=-\frac{{\lambda}_{20}(\partial_t{u})^{(0,1)}-{\lambda}_{11}(\partial_t{u})^{(1,0)}}{{\lambda}_{20}{\lambda}_{02}-{\lambda}_{11}^2} \Big\rvert_{\kappa_\phi,\kappa_\chi}\,,
\end{align}
and
\begin{align}
\hspace{-0.1cm}\partial_t{{\lambda}}_{ij}=(\partial_t{u})^{(i,j)} +u^{(i+1,j)}\partial_t{\kappa}_{\phi}+u^{(i,j+1)}\partial_t{\kappa}_{\chi}\Big\rvert_{\kappa_\phi,\kappa_\chi}\,,\nonumber
\end{align}
where the field invariants $\rho_j$ are understood to be evaluated at
their scale dependent expansion points $\kappa_j$.


\section{Calculation of Explicit Loop Flows}
\label{app:Exflow}

\subsubsection*{Explicit Loop flows I}
\label{app:Exflow1}

\noindent Starting with Eqs.~\eq{dulinsup} and \eq{dulinsupB}, we can calculate a flow which translates continuously from one regulator to another as long as both regulators are finite. These equations can be easily extended to an $O(N)$ model
\begin{align}
\frac{\text{d}}{\text{d}s} u = &J(u'+2\rho\, u'')+(N-1)J(u')\nonumber\\
&+\frac{\partial_s k(s)}{k(s)}\big(-d\,u+(d-2)\rho\, u'\big)\,.
\end{align}
A commonly used regulator is the sharp regulator $r_{\text{sharp}}(y)=c/y \ \theta(1-y)|_{c\rightarrow \infty}$, which is infinite in [0,1]. In order to interpolate between $r_{\text{sharp}}$ and other regulators in a continuous manner we need to extend our calculations.
Here, we interpolate between $r_{\text{sharp}}$ and $r_{\text{L}}$ using an interpolation which shifts the cutoff scale in $r_{\text{sharp}}$ by a factor $a(s)\in [0,1]$.
\begin{align}
r_s(y)&=r_{\text{L}}(y)+r_{\text{sharp}}\big(y/a(s)^2\big) \, ,
\end{align}
Hence, $a(s)\rightarrow0$ causes $r_{\text{sharp}}$ to vanish. On the other hand, if $a(s)=1$, then $r_{\text{sharp}}$ causes the regulator to diverge on $[0,1]$ such that there is no residual influence of $r_{\text{L}}$.
The threshold function can be decomposed into two parts
\begin{align}
J(\omega)&=v_d\int_0^\infty y^{\frac{d}{2}}\textrm{d}y \frac{\frac{d}{ds}r_s(y)}{y(1+ r_s(y))+\omega}
\nonumber\\
&=J^{\text{A}}(\omega)+J^{\text{B}}(\omega)\,,\label{th2}
\end{align}
where $J^{\text{A}}$ contains the regulator derivative from
$r_{flat}$ such that it can be inferred from \eq{thflat}
\begin{align}
J^{\text{A}}(\omega)&=\frac{k'(s)}{k(s)} (1-a(s)^d) I^{\text{(L)}}(\omega)\nonumber\\
&=v_d\frac{k'(s)}{k(s)} \frac{4}{d}\frac{1}{1+w}(1-a(s)^d)\,.
\end{align}
Similarly, $J^{\text{B}}$ corresponds to the regulator derivative of
$r_{sharp}$ and can be calculated by inserting \eq{thsharp}
\begin{align}
J^{\text{B}}(\omega)&=\Big(\frac{a'(s)}{a(s)}+\frac{k'(s)}{k(s)}\Big)a(s)^{d-2}I^{\text{(sharp)}}(\omega)\nonumber\\
&=-2v_d\Big(\frac{a'(s)}{a(s)}+\frac{k'(s)}{k(s)}\Big)a(s)^{d-2}\log(1+\omega) \, .
\end{align}


\subsubsection*{Explicit Loop flows II}
\label{app:Exflow2}
\begin{figure}[t!]
  \centering
  \includegraphics[width=.95\columnwidth]{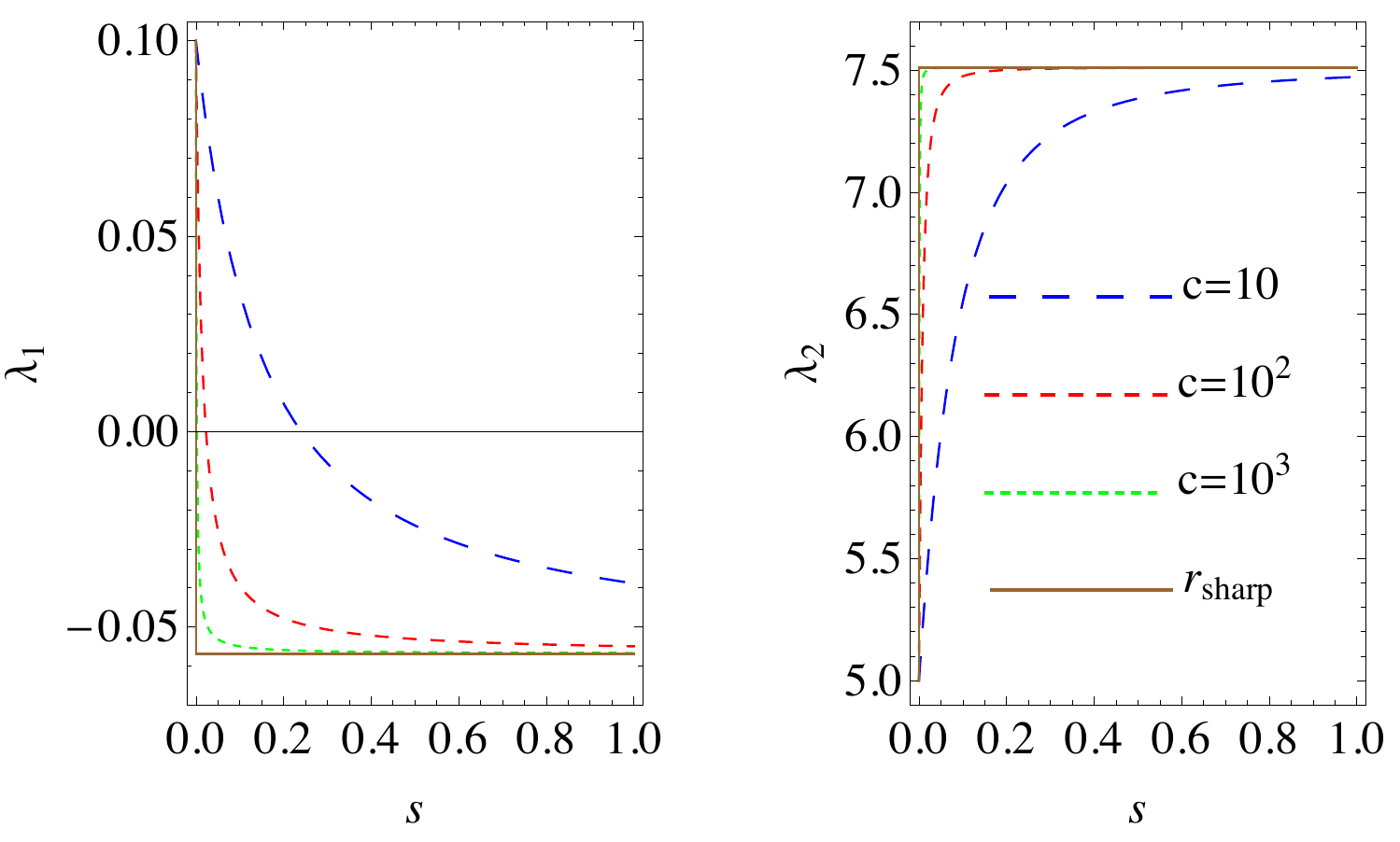}
  \caption{Change of the initial conditions for different choices of
    interpolations between the flat regulator and the step like
    regulator, for various $c$ and the sharp regulator.}
\label{icond}
\end{figure}

\noindent In case we insist on a linear superposition between
$r_{\text{sharp}}$ and other finite regulators, the solution of the
flow equation will show a discontinuity. As soon as we allow for a
small contribution from $r_{\text{sharp}}$, it will dominate over all
finite regulators because it is infinite in the region $[0,1]$. This
discontinuity can be seen in Figs.~\ref{1loop} and \ref{icond} for the
couplings $\lambda_1$ and $\lambda_2$. In order to calculate the
magnitude of this discontinuity, we start at the flow equation at
fixed $k$ for an $O(1)$-model, \Eq{dulinsup}
\begin{align}
\frac{\text{d}}{\text{d}s} u\rvert_{k=\text{const}} = J(\omega) \, , \  \omega=u'+2\rho\, u'' \, ,
\end{align}
which we want to solve from $s=0$ to $s=1$. Since the only change
occurs at $s=0$, we can simply denote the magnitude of the
discontinuity $\Delta u=u(s=1)-u(s=0)$.
The threshold function 
\begin{align}
&J(\omega)=v_d\int_0^\infty y^{\frac{d}{2}}\textrm{d}y \frac{\partial_s r_s(y)}{y(1+ r_s(y))+\omega}\,,
\end{align}
only depends on the change of shape $\partial_s r_s(y)$, but not on
the scale change, since we are keeping $k$ fixed. Inserting a linear
superposition between $r_{\text{sharp}}$ and $r_{\text{L}}$
\begin{align}
r_s(y)&=(1-s)r_{\text{L}}(y)+s\ r_{\text{sharp}}(y)\,,
\end{align}
evaluates to
\begin{align}
  J(w)=v_d\int_0^1
  y^{\frac{d}{2}-1}\textrm{d}y\,\frac{c(1+1/c-y/c)}{1+sc(1-1/c+y/c)+w}\Big|_{c\rightarrow
    \infty}\nonumber\,,
\end{align}
We now shift the flow variable from $s\in[0,1]$ to $\bar
s=s\,c\in[0,c]$ and take the limit $c\rightarrow \infty$.
The new modified flow equation, reading
\begin{align}
&\frac{\text{d}}{\text{d}\bar s} u\rvert_{k=\text{const}} =\frac{2 v_d}{d} \frac{1}{\bar s + 1+w}\,,
\end{align}
must be solved from $\bar s=0$ to $\bar s =\infty$. 
This equation leads to a logarithmic divergence if $\bar s\rightarrow
\infty$. However, the divergent part is just a constant shift of the
effective potential which can be removed by subtracting it
\begin{align}
  \frac{\text{d}}{\text{d}\bar s} u\rvert_{k=\text{const}} &=-\frac{2
    v_d}{d}\frac{1}{\bar s + 1} \frac{\omega}{\bar s + 1+\omega}\,.
\end{align}
Our construction ensures that the discontinuity is expressed as
$\Delta u= u(\bar s=\infty)-u(\bar s = 0)$ which can be evaluated in a
continuous flow equation.
\begin{figure}[t!]
  \centering
  \includegraphics[width=0.9\columnwidth]{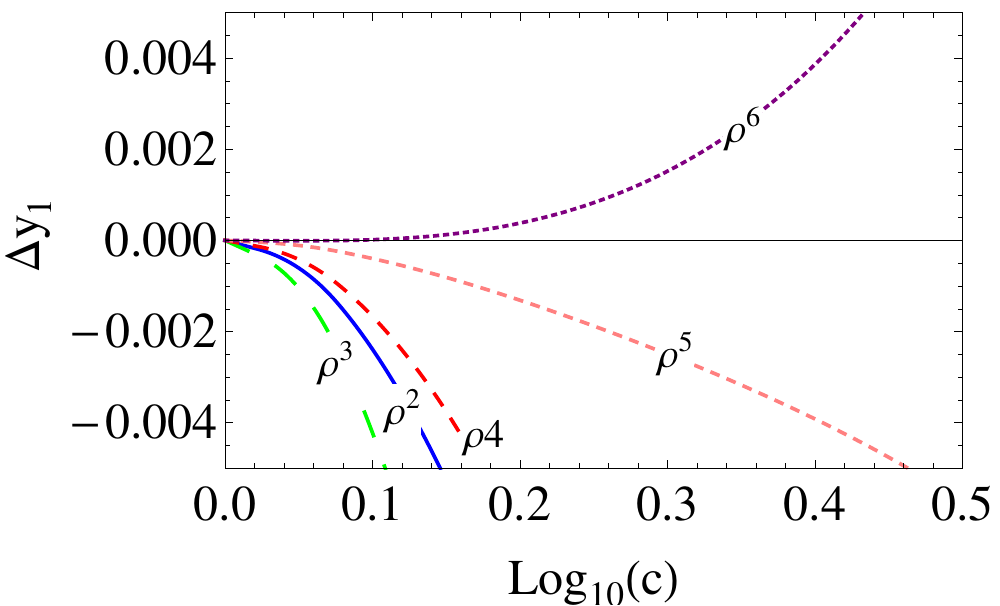}
  \caption{Deviation from the critical exponent $y_1$ using separated
    cutoff scales in the $O(2)\oplus O(2)$ model in LPA from order
    $\rho^2$ to order $\rho^6$.}
\label{CritExpStabi02}
\end{figure}
\begin{figure}[t!]
  \centering
  \includegraphics[width=0.8\columnwidth]{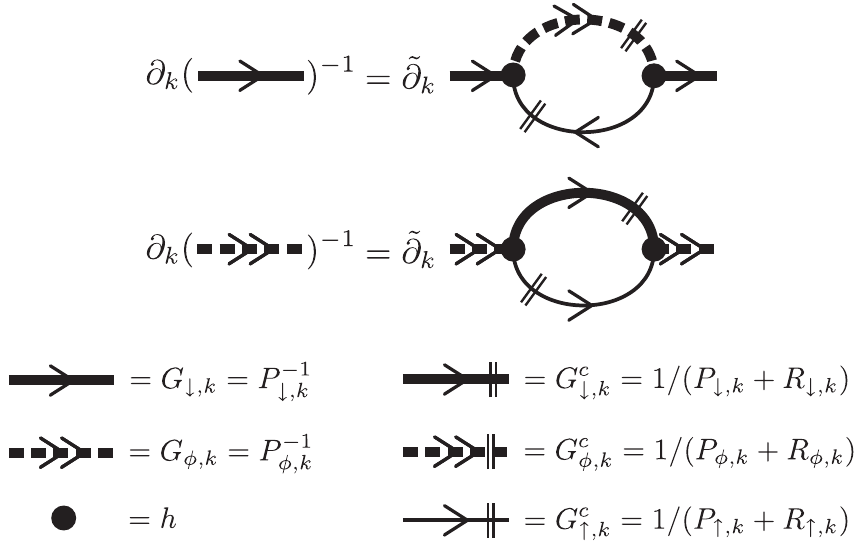}
  \caption{Feynman diagrams for the flow of the dimer and impurity
    inverse propagators.}
\label{fig.pol.diagrams}
\end{figure}


\section{Two-Field-Model \& Critical Exponents}\label{app:2fieldexp}

\noindent In Fig.~\ref{CritExpStabi02}, we show a variant of Fig.~\ref{CritExpStababs},
exhibiting the deviation of the critical exponent $y_1$ in the
$O(2)\oplus O(2)$ model from its value at $c=1$ without taking the
absolute value and the weighting factor from \eq{eq:absweight}.


\section{Flows for the Polaron Problem}\label{app:polaron}

\noindent The flow equations, graphically represented in Fig.~\ref{fig.pol.diagrams}, are given by
 \begin{align}
  \partial_k P_{\downarrow,k}(Q)&=
  h^2\tilde\partial_k\int_P G_{\phi,k}^c(P) G_{\uparrow,k}^c(P+Q)\notag\\[2ex]
  \partial_k P_{\phi,k}(Q)&=
  -h^2\tilde\partial_k\int_P G_{\downarrow,k}^c(P) G_{\uparrow,k}^c(Q-P),
  \label{pol.genflow}
\end{align}
where $ P\equiv (\omega, \bf{p})$. The  flowing inverse propagators $P_k \equiv G_k^{-1}$ on the left-hand side are defined without the regulators, while the regulated propagators $G_k^c$ are given by
\begin{align}
  \label{pol.defP}
  G_k & \equiv 1/P_k &
  G_k^c & \equiv 1/(P_k + R_k),
\end{align}
and $\tilde\partial_k$ implies that the derivative acts only on the regulator term $R_k$ inside the cutoff propagators $G_k^c$.



\end{document}